\documentclass[aps,prl,preprint,superscriptaddress,floatfix]{revtex4-2}

\usepackage[english]{babel}
\usepackage[dvips]{graphicx}
\usepackage{graphics}
\usepackage{amsfonts}
\usepackage{amsmath}
\usepackage{amssymb}
\usepackage{exscale}
\usepackage{eufrak}
\usepackage{afterpage}
\usepackage{upgreek}
\usepackage{color}	
\usepackage{braket}	
\usepackage{multirow}	
\usepackage{enumitem}
\usepackage{balance}

\usepackage{miller}	

\usepackage[normalem]{ulem}    

\begin{document}

\title{Imaging the itinerant-to-localized transmutation of electrons 
	  across the metal-to-insulator transition in V$_2$O$_3$
	   }
\author{Maximilian~Thees}
\affiliation{Universit\'e Paris-Saclay, CNRS,  Institut des Sciences Mol\'eculaires d'Orsay, 
			91405, Orsay, France}

\author{Min-Han~Lee}
\affiliation{Department of Physics and Center for Advanced Nanoscience, University of California,
			 San Diego, La Jolla, California 92093, USA}

\author{Rosa~Luca~Bouwmeester}
\affiliation{Faculty of Science and Technology and MESA+ Institute for Nanotechnology, 
			University of Twente, 7500 AE Enschede, The Netherlands} 			

\author{Pedro~H.~Rezende-Gon\c calves}
\affiliation{Universit\'e Paris-Saclay, CNRS,  Institut des Sciences Mol\'eculaires d'Orsay, 
			91405, Orsay, France}
\affiliation{Departamento de F\'isica, Universidade Federal de Minas Gerais, 
			 Av. Pres. Antonio Carlos, 6627, Belo Horizonte, Brazil}

\author{Emma~David}
\affiliation{Universit\'e Paris-Saclay, CNRS,  Institut des Sciences Mol\'eculaires d'Orsay, 
			91405, Orsay, France}
			
\author{Alexandre~Zimmers}
\affiliation{LPEM, ESPCI Paris, PSL Research University, CNRS, Sorbonne Universit\'e, 75005 Paris, France} 
			
\author{Franck~Fortuna}
\affiliation{Universit\'e Paris-Saclay, CNRS,  Institut des Sciences Mol\'eculaires d'Orsay, 
			91405, Orsay, France} 
			
\author{Emmanouil~Frantzeskakis}
\affiliation{Universit\'e Paris-Saclay, CNRS,  Institut des Sciences Mol\'eculaires d'Orsay, 
			91405, Orsay, France} 

\author{Nicolas~M.Vargas}
\affiliation{Department of Physics and Center for Advanced Nanoscience, University of California,
			 San Diego, La Jolla, California 92093, USA}

\author{Yoav~Kalcheim}
\affiliation{Department of Physics and Center for Advanced Nanoscience, University of California,
			 San Diego, La Jolla, California 92093, USA}
\affiliation{6Department of Materials Science and Engineering, 
			 Technion - Israel Institue of Technology, Haifa 3200003, Israel}
			
\author{Patrick~Le~F\`evre}
\affiliation{Synchrotron SOLEIL, L'Orme des Merisiers, Saint-Aubin-BP48, 91192 Gif-sur-Yvette, France}

\author{Koji~Horiba}
\affiliation{Photon Factory, Institute of Materials Structure Science, 
High Energy Accelerator Research Organization (KEK), 1-1 Oho, Tsukuba 305-0801, Japan}

\author{Hiroshi~Kumigashira}
\affiliation{Institute of Multidisciplinary Research for Advanced Materials (IMRAM), 
			Tohoku University, Sendai, 980-8577, Japan}
\affiliation{Photon Factory, Institute of Materials Structure Science, 
			High Energy Accelerator Research Organization (KEK), 1-1 Oho, Tsukuba 305-0801, Japan}

\author{Silke~Biermann}
\affiliation{CPHT, CNRS, Ecole Polytechnique, Institut Polytechnique de Paris, F-91128 Palaiseau, France}
\affiliation{Coll\`ege de France, 11 place Marcelin Berthelot, 75005 Paris, France}
\affiliation{Department of Physics, Division of Mathematical Physics, Lund University, 
			Professorsgatan 1, 22363 Lund, Sweden}

\author{Juan~Trastoy}
\affiliation{Unit\'e Mixte de Physique, CNRS, Thales, Universit\'e Paris-Sud, Universit\'e Paris-Saclay,
			 91767 Palaiseau, France}

\author{Marcelo~J.~\surname{Rozenberg}}
\affiliation{Universit\'e Paris-Saclay, CNRS, Laboratoire de Physique des Solides, Orsay 91405, France}

\author{Ivan~K.~Schuller}
\affiliation{Department of Physics and Center for Advanced Nanoscience, University of California,
			 San Diego, La Jolla, California 92093, USA}

\author{Andr\'es~F.~Santander-Syro}
\email{andres.santander-syro@universite-paris-saclay.fr}
\affiliation{Universit\'e Paris-Saclay, CNRS,  Institut des Sciences Mol\'eculaires d'Orsay, 
			91405, Orsay, France} 


\begin{abstract}
  \textbf{
  In solids, strong repulsion between electrons can inhibit their movement 
  and result in a ``Mott" metal-to-insulator transition (MIT), 
  a fundamental phenomenon whose understanding has remained a challenge for over 50 years. 
  A key issue is how the wave-like itinerant electrons change into a localized-like state 
  due to increased interactions. However, observing the MIT 
  in terms of the energy- and momentum-resolved electronic structure of the system, 
  the only direct way to probe both itinerant and localized states, has been elusive. 
  Here we show, using angle-resolved photoemission spectroscopy (ARPES), 
  that in V$_2$O$_3$ the temperature-induced MIT is characterized 
  by the progressive disappearance of its itinerant conduction band, 
  without any change in its energy-momentum dispersion, 
  and the simultaneous shift to larger binding energies of a quasi-localized state 
  initially located near the Fermi level.
  }
\end{abstract}
%
\maketitle

\begin{figure}[p!]
	\centering
        \includegraphics[clip, width=0.9\linewidth]{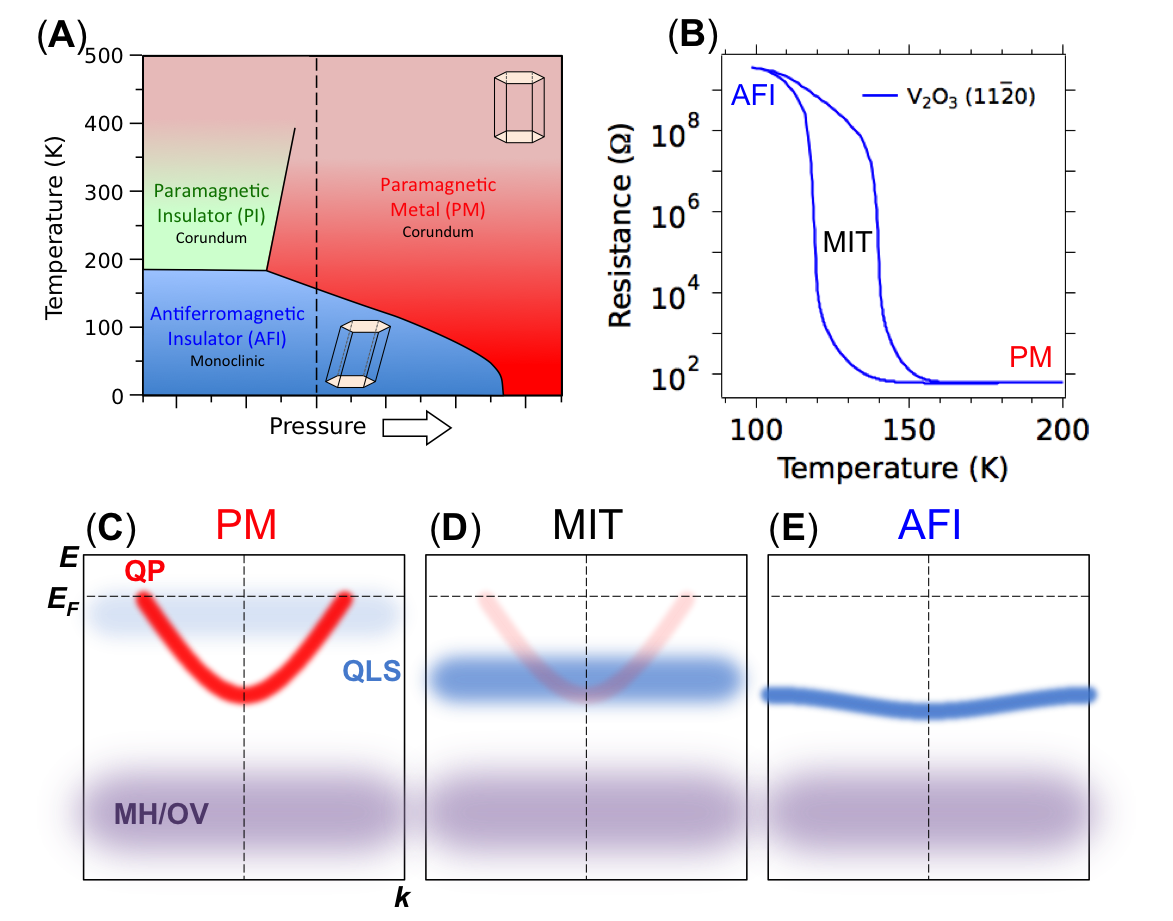}
    \caption{\label{Fig:V2O3_General}
    		 \footnotesize{
    		 \textbf{V$_2$O$_3$: schematic phase diagram and electronic-structure changes.}
        	 (A)~Generic temperature-pressure phase diagram of bulk V$_2$O$_3$~\cite{McWhan-1971}.
        	 (B)~Electrical resistance of a V$_2$O$_3$/Al$_2$O$_3$\hkl(11-20) thin film
        	 studied in this work, showing the paramagnetic metal (PM),
        	 antiferromagnetic insulator (AFI), and the coexistence region across the MIT.        	 
        	 (C,~D,~E)~Schematic representation of the near-$E_F$ electronic-structure
        	 evolution observed in this work. In the PM phase, a QP band and a QLS are observed 
        	 near $E_F$. In the coexistence region, the QP looses spectral weight 
        	 without changes in dispersion,
        	 while the QLS shifts down in energy and gains spectral weight. 
        	 In the AFI phase, only the QLS (showing a slight dispersion) remains, 
        	 resulting in a gap below $E_F$.
        	 }
         }
\end{figure}

According to the quantum-mechanical band theory of solids, 
in insulators the highest occupied band is totally filled, 
while in metals it is partially filled~\cite{Aschroft-Mermin-Book}. 
Thus, as temperature cannot change the number of electrons in a solid, 
it should not change either its intrinsic nature, i.e. metallic or insulating. 
Enter V$_2$O$_3$: the formal configuration of the vanadium ion would be V$^{+3}[3d^2]$, 
hence this oxide should be a metal. 
However, bulk V$_2$O$_3$ shows a first-order metal-to-insulator transition (MIT) 
when cooling below $T_{\textrm{MIT}} \approx 160$~K, with an abrupt resistivity change 
of over six orders of magnitude~\cite{Fulde-2012,Mott-1968,McWhan-1969,McWhan-1971,Mott-1990,
Imada-1998,Dobrosavljevic-2012} --see Figs.~\ref{Fig:V2O3_General}(A,~B).
Microscopically, the MIT is characterized by gap of about $750$~meV opening 
in the optical conductivity~\cite{Rozenberg-1995,Rozenberg-1996},
and is accompanied by corundum-to-monoclinic and paramagnetic to antiferromagnetic 
transitions~\cite{McWhan-1969,Mott-1990,Moon-1970,Bao-1997,Imada-1998},
as illustrated in Fig.~\ref{Fig:V2O3_General}(A). 
In fact, due to the relatively localized character of the $3d$-orbitals,
the partially-filled bands of V$_2$O$_3$ are prone to 
strong electronic correlations, neglected in band theory~\cite{Fulde-2012}. 
V$_2$O$_3$ is thus considered an archetypal system for the Mott MIT, 
one of the most fundamental manifestations of electron correlations,
also observed in several other materials. 
However, after 50 years of research, the microscopic processes accompanying the Mott MIT, 
including the roles played by the electronic, magnetic and structural degrees of freedom, 
are still controversial issues~\cite{McWhan-1969,Mott-1990,Bao-1997,
Imada-1998,Dobrosavljevic-2012,Rozenberg-1995,Anisimov-2005,Poteryaev-2007,
Trastoy-2020,Kalcheim-2020}. 
A major reason is that an experimental imaging of the momentum-resolved changes 
in the electronic structure of any Mott system 
across the thermally induced MIT is still missing.
Previous photoemission experiments
studied the momentum-integrated density of states 
of V$_2$O$_3$, identifying a quasi-particle (QP) peak at the Fermi level ($E_F$) 
in the metallic phase~\cite{Mo-2003} that disappears in the insulating state~\cite{Mo-2006}, 
and a broad feature at a binding energy $E - E_F \approx -1.1$~eV~\cite{Shin-1990,Smith-1994}, 
assigned to the lower Mott-Hubbard (MH) band~\cite{Mo-2003,Mo-2006}. 
The coexistence of the QP peak and the MH band is the hallmark of the correlated state, 
as predicted by dynamical mean-field theory some 30 years ago~\cite{Georges-1992}.
More recently, ARPES experiments in the metallic phase of V$_2$O$_3$ 
single crystals~\cite{Lo-Vecchio-2016} showed the existence 
of an electron-like QP band around the zone center ($\Gamma$ point), 
dispersing down to about $-400$~meV over Fermi momenta $2 k_F \approx 1$~\AA$^{-1}$ 
along the $\Gamma \textrm{Z}$ direction, and suggested the presence of a non-dispersive component 
of spectral weight in the metallic QP energy region.
However, several technical challenges have hindered the realization 
of momentum-dependent photoemission studies of the MIT in V$_2$O$_3$. 
For instance, V$_2$O$_3$ crystals are extremely hard to cleave in order to expose
a clean crystalline surface. Moreover, they become highly insulating below $T_{\textrm{MIT}}$,
thus strongly charging upon electron emission,
and break apart into pieces, due to the structural transition. 
The main physical issue at stake is at the core of the strong correlations problem, 
namely how electrons transmute from wave-like objects in the metallic phase 
to localized particles in the insulating one. 
In this work, we were able to directly address this question experimentally, 
and we provide answers to a variety of key questions:  
the evolution of the Mott gap, QP dispersions, effective masses, 
orbital character and relative spectral weights of the QP and Mott-Hubbard bands, 
and an understanding, from the viewpoint of electronic structure, 
of the hysteresis cycle observed in the MIT.

Using angle-resolved photoemission spectroscopy (ARPES), 
we studied high-quality crystalline thin films of 
V$_2$O$_3$ grown on Al$_2$O$_3$ substrates~\cite{Valmianski-2017,Trastoy-2018,Kalcheim-2019}. 
Thanks to the anchoring imposed by the latter, the crystal integrity of the films 
is not affected by stress due to the structural transition. 
This allowed us to measure, for the first time, 
the effects of the MIT on the \emph{momentum-resolved} 
spectral function of the system.
As schematized in Figs.~\ref{Fig:V2O3_General}(C-E),
we found that the opening of the Mott gap at $E_F$ 
in energy-momentum space happens \emph{abruptly}
following a gradual spectral-weight transfer:
as temperature decreases in the MIT regime, there is a progressive decrease in spectral weight 
of an itinerant, i.e. dispersive quasi-particle (QP) conduction band,
without noticeable changes in its dispersion and effective mass.
This is accompanied by a shift, and  increase in spectral weight,
of a quasi-localized state (QLS), which goes from an energy close to $E_F$ in the metallic state
to an energy close to the bottom of the vanishing dispersive band in the insulating state.
Only when the dispersive state crossing the Fermi level has vanished
a complete gap of about $700$~meV with respect to $E_F$ is observed, 
associated to the final energy position of the QLS.
Furthermore, the spectral weight of the above-mentioned near-$E_F$ features 
shows a clear thermal hysteresis that tracks the one observed in macroscopic transport data.
Another non-dispersive state at lower binding energy, 
associated to the lower Mott-Hubbard (MH) band, 
and possibly also containing a contribution from to oxygen vacancy (OV) states (see later),
does not show an appreciable variation with temperature.

\begin{figure*}[p!]
	\centering
        \includegraphics[clip, width=0.8\linewidth]{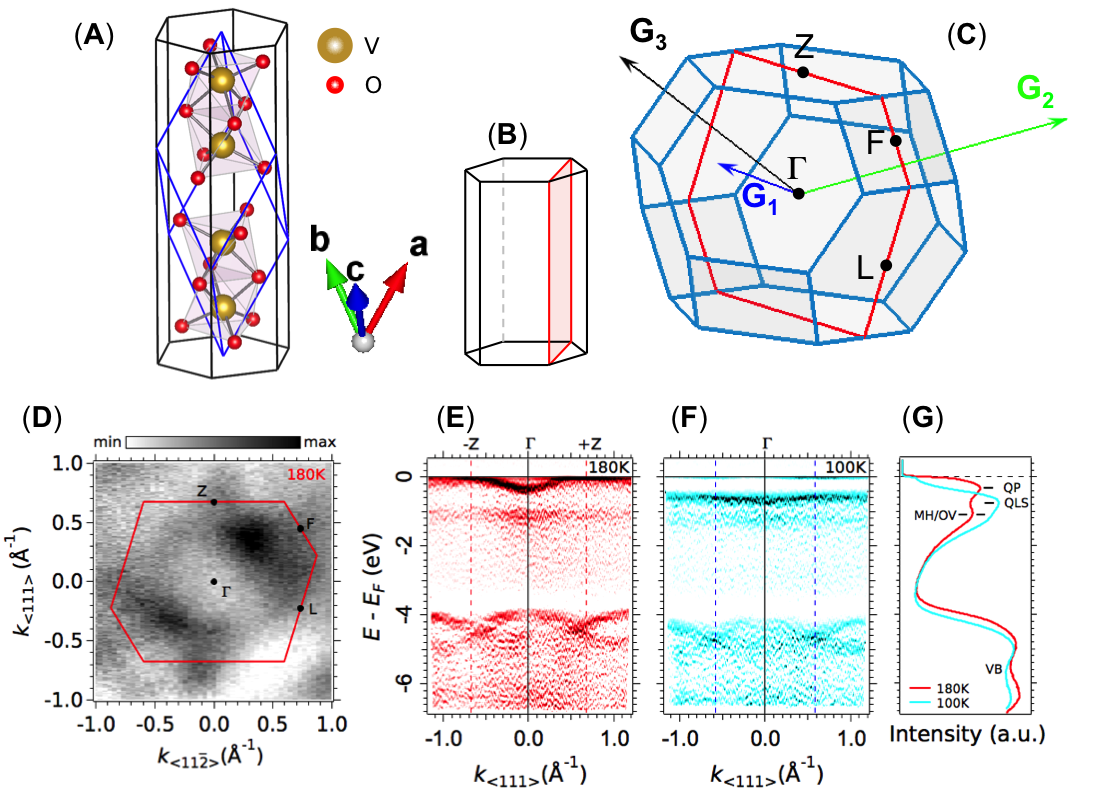}
    \caption{\label{Fig:V2O3_ARPES_Gral}
    		 \footnotesize{
    		 \textbf{V$_2$O$_3$: Crystal structure and electronic gap opening across the MIT.}
        	 (A)~Rhombohedral crystal structure of V$_2$O$_3$ (blue polyhedron)
        	 in the high-temperature metallic state.  
        	 The non-primitive hexagonal cell (black polyhedron) is also shown.
        	 (B)~Representation of the \hkl(11-20) plane (in red) measured in this work
        	 inside the non-primitive hexagonal cell.
        	 (C)~Rhombohedral Brillouin zone, showing the primitive vectors of the
        	 reciprocal lattice. The \hkl(-110) plane, corresponding to
        	 a \hkl(11-20) plane in hexagonal coordinates, is shown in red.
        	 (D)~Fermi-surface map in the \hkl<111>-\hkl<11-2> plane
        	 of a thin film of V$_2$O$_3$/Al$_2$O$_3$\hkl(11-20)
        	 in the metallic state at 180~K. 
			 The rhombohedral Brillouin zone edges are marked in red.
			 Associated high-symmetry points are indicated.
			 (E,~F)~Energy-momentum maps (2D curvature)
			 along $\Gamma \textrm{Z}$ in the metallic ($180$~K) and insulating ($100$~K) states, 
			 respectively, showing four bands: 
			 an electron-like QP band at the Fermi-level, visible at $180$~K; 
			 a weakly dispersive QLS at $E - E_F \approx -700$~meV,
			 best observed at $100$~K;
			 a broad and weakly dispersing MH/OV band 
			 around $E - E_F \approx -1.1$~eV, seen at both temperatures;
			 and the VB of oxygen $p$-states 
			 extending from $E - E_F \approx -4$~eV downwards,
			 also present at all temperatures.
			 The rhombohedral Brillouin zone edges ($\pm \textrm{Z}$) at $180$~K,
			 and the monoclinic zone edges at $100$~K, 
			 are indicated by red and blue dashed lines, respectively.
			 (G)~Momentum-integrated raw ARPES intensities from (E) and (F),
			 showing the QP, QLS, MH/OV and VB states.
			 All data were measured at a photon energy of $86$~eV,
			 corresponding to a bulk $\Gamma$ point in the \hkl<-110> direction,
			 using linear horizontal polarized light.
			 The color hues in panels (D,~E,F), as in the rest of this paper, 
			 indicate the ARPES intensity between the minimum (min) and maximum (max) 
			 detected signal.
			 See the Methods for a detailed description of the 
			 ARPES measurements and curvature analyses.
        	 }
        }
\end{figure*}

Fig.~\ref{Fig:V2O3_ARPES_Gral}(A) shows the crystal structure of V$_2$O$_3$ 
both in its primitive high-temperature rhombohedral unit cell (blue) 
and in the associated conventional hexagonal cell (black).
In rhombohedral coordinates of the primitive cell, 
we write as \hkl(hkl) the orientation of crystallographic planes,
and as \hkl<hkl> directions in reciprocal space.
In hexagonal coordinates of the conventional cell, we employ the four Miller indices notation,
writing planes and directions respectively as \hkl(hkil) and \hkl<hkil> (with $i=-h-k$).  

In this work we measured V$_2$O$_3$/Al$_2$O$_3$\hkl(11-20) thin films,
whose surface, schematically shown in Fig.~\ref{Fig:V2O3_ARPES_Gral}(B), 
is perpendicular to the basal plane of the hexagonal cell.
Complementary data on V$_2$O$_3$/Al$_2$O$_3$\hkl(01-12) films
are presented in the Supplementary Information.
The surface of the V$_2$O$_3$ films was cleaned \emph{in-situ} using protocols 
previously developed for the investigation of two-dimensional (2D) electron gases 
in oxides~\cite{Roedel-2016,Backes-2016,Roedel-2017,Loemker-2017,Roedel-2018}.
The cleaned surfaces showed well-defined low-energy electron diffraction (LEED) 
patterns (Supplementary Information).
The cleaning process slightly lowers the onset temperature of the MIT 
and decreases the change in resistance between the insulating and metallic states,
possibly due to the formation of 
oxygen vacancies~\cite{Santander-2011,Santander-2012,Roedel-2016,Backes-2016},
but does not affect the stoichiometry of the film 
nor the overall physical changes across the transition 
(Supplementary Information).
See the Methods for technical details about our 
thin-film growth, characterization, ultra-high-vacuum (UHV) annealing
and ARPES measurements.

Fig.~\ref{Fig:V2O3_ARPES_Gral}(C) shows the 3D rhombohedral Brillouin zone of V$_2$O$_3$
in its metallic phase, together with the 2D plane through $\Gamma$ 
parallel to the surface of our films. 
For simplicity in notation, everywhere in this work the ARPES data 
will be referred to directions in this particular Brillouin zone,
both in the metallic and insulating phases.
When relevant, the Brillouin zone edges of the monoclinic insulating structure 
will be indicated. The sample's surface orientation will be specified 
using hexagonal coordinates, as commonly done in the thin-film literature. 
The Supplementary Information discusses further 
the rhombohedral and monoclinic Brillouin zones in relation to our ARPES data 
in the metallic and insulating phases. 

We now present the ARPES data across the MIT.
Fig.~\ref{Fig:V2O3_ARPES_Gral}(D) shows the Fermi surface map in the \hkl(-110) 
plane of a V$_2$O$_3$/Al$_2$O$_3$\hkl(11-20) thin film 
measured in the metallic state at $T=180$K.
One observes a large Fermi sheet around the center of the Brillouin zone ($\Gamma$ point).
Photon-energy-dependent ARPES data presented in the 
Supplementary Information (Fig.~\ref{Fig:V2O3_ARPES_FermiSurfaces}) 
show that the Fermi surface disperses in the momentum direction 
perpendicular to the sample surface, demonstrating that the measured states 
are intrinsic to the bulk three-dimensional electronic structure of the material. 
Fig.~\ref{Fig:V2O3_ARPES_Gral}(E) presents the corresponding energy-momentum ARPES map
along $k_{<111>}$, corresponding to the $\Gamma \textrm{Z}$ direction
in the rhombohedral metallic phase. 
The most evident features are an electron-like QP band
crossing the Fermi level ($E_F=0$) and dispersing down to
about $-400$~meV~\cite{Mo-2006,Lo-Vecchio-2016}, 
together with a non-dispersive state around an energy $E=-1.1$~eV, 
assigned to the lower MH band~\cite{Mo-2006}, 
and the valence band (VB) of oxygen $p$-states below about $E=-4$~eV.
All these features are in excellent agreement with previous photoemission 
and ARPES measurements in the metallic state of single crystals~\cite{Mo-2006,Lo-Vecchio-2016}.
The clear dispersion of the QP and valence bands is, moreover, an experimental proof 
of the crystalline quality of the thin film surface.
Note also that the Mott-Hubbard band has most of its spectral weight concentrated 
at momenta around $\Gamma$, below the QP band bottom, 
similarly to what has been previously seen in other 
correlated-electron metals~\cite{Takizawa-2009,Backes-2016}.
Part of the non-dispersive spectral weight present at the same energy as the MH band,
also observed in previous ARPES works on V$_2$O$_3$~\cite{Lo-Vecchio-2016},
might arise from localized states associated to 
the creation of oxygen vacancies during the annealing process
and/or UV irradiation during experiments.
Such vacancy states are commonly found at about the same binding energy,
namely  $E - E_F \approx -1$~to~$-1.5$~eV, in virtually all 
transition metal oxides~\cite{Roedel-2016,Roedel-2017,Loemker-2017,Santander-2011,Santander-2012},
including the correlated metal SrVO$_3$, where they indeed superpose
with the MH band~\cite{Backes-2016}.

Thanks to our thin films that preserve their crystal integrity upon cooling,
we can now measure the momentum-resolved spectra in the insulating phase.
Fig.~\ref{Fig:V2O3_ARPES_Gral}(F) shows the energy-momentum ARPES map 
along $k_{<111>}$ at $T=100$~K,
in the insulating state of the V$_2$O$_3$/Al$_2$O$_3$\hkl(11-20) thin film.
While the VB and the MH/OV bands remain essentially unchanged,
the states near $E_F$ show a dramatic reconstruction:
instead of the strongly dispersive QP state, 
one observes now a weakly dispersing, previously unreported quasi-localized state (QLS), 
at $E - E_F \approx -700$~meV,
\emph{different} from the MH/OV band.
The energy of this QLS gives thus a lower bound to the Mott gap.
This agrees well with the gap of about $750-800$~meV
observed in previous optical conductivity studies 
on V$_2$O$_3$ single crystals~\cite{Rozenberg-1995,Rozenberg-1996}.
It also agrees with our own infrared measurements on the same thin films
used for our ARPES experiments, which show a strong decrease in reflectivity 
below about $800-900$~meV in the insulating phase 
(Supplementary Information, Fig.~\ref{Fig:V2O3_IR_Annealing}).
Note that optical measurements yield the true energy gap 
between the highest fully occupied state, hence the QLS, 
and the first unoccupied state above $E_F$, not accessible to ARPES.
The QP, QLS, MH/OV and VB states can also be seen in the momentum-integrated
ARPES intensities, Fig.~\ref{Fig:V2O3_ARPES_Gral}(G).
In fact, as we will see next, the QP and QLS states are \emph{a priori} of different nature:
the QLS is also present in the metallic phase
at energies near $E_F$, where it coexists with the QP state
--as hinted by previous ARPES work on the metallic phase of 
V$_2$O$_3$ single crystals~\cite{Lo-Vecchio-2016}. 
But as the system becomes insulating, the QLS shifts down in energy
and increases in intensity, while the QP state gradually loses its spectral weight.

\begin{figure*}[p!]
	\centering
        \includegraphics[clip, width=\linewidth]{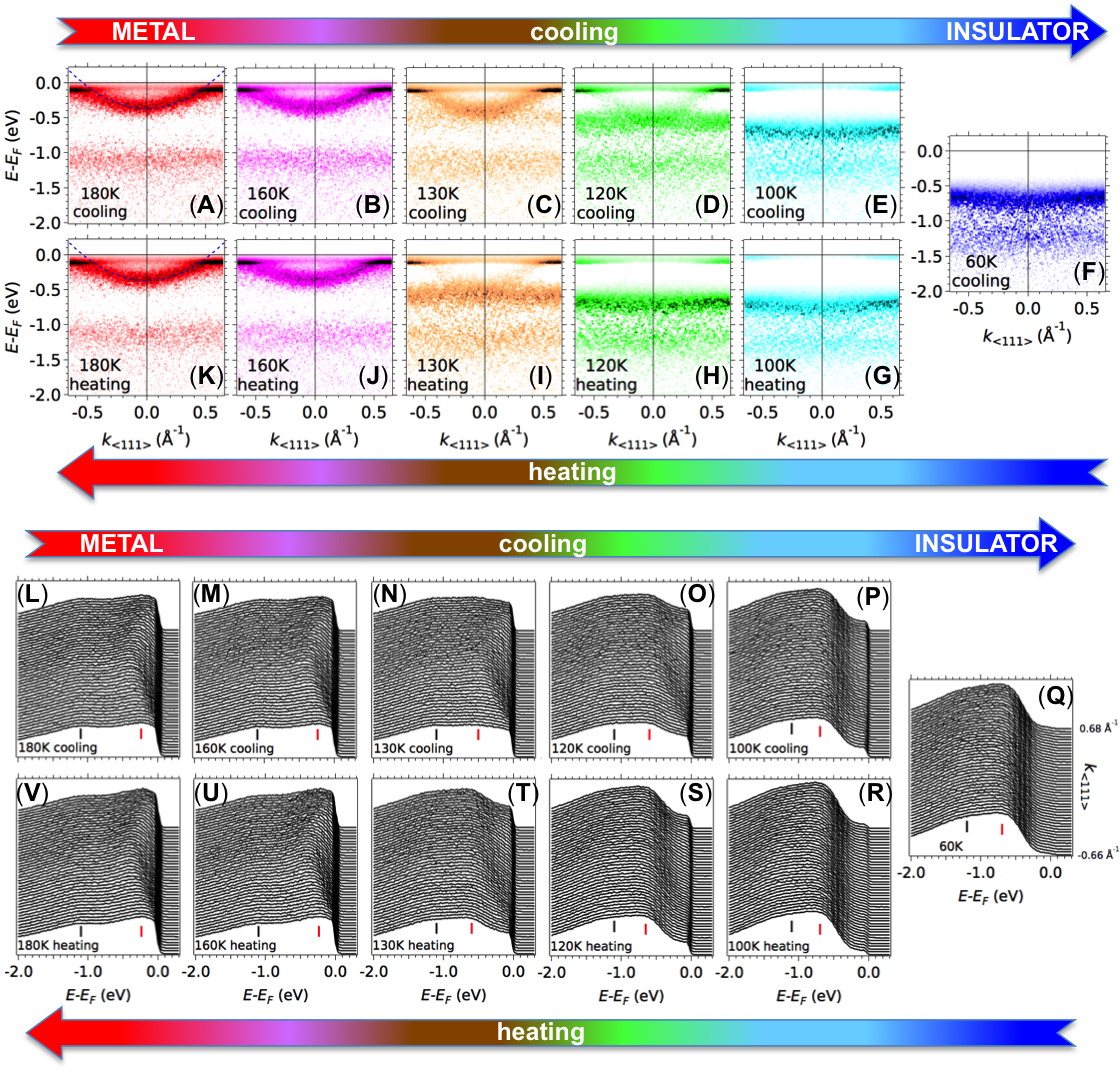}
    \caption{\label{Fig:V2O3_ARPES_Tdep}
    		 \footnotesize{
    		 \textbf{Reconstruction and hysteresis of the electronic structure across the MIT.}
        	 (A-F)~Evolution of the ARPES energy-momentum spectra near $E_F$ 
        	 (2D curvature of intensity maps, see Methods)
        	 when cooling from $180$~K (metallic state) to $60$~K (insulating state) 
        	 in a V$_2$O$_3$\hkl(11-20) thin film. 
        	 The sharp pile-up of intensity at $E_F$ 
        	 is a spurious effect of the 2D curvature analysis on the Fermi-Dirac cutoff. 
        	 (G-K)~Corresponding spectra when heating back to $180$K.
        	 The blue dashed parabolas in (A) and (K) represent a quasi-free electron band 
        	 of effective mass $m^{\star} = 3.5 m_e$, assigned to the QP band.        	 
        	 (L-V)~Raw data associated to (A-K). Red and black markers indicate the positions
        	 of the QLS and MH/OV bands, respectively.
        	 }
        }
\end{figure*}

Fig.~\ref{Fig:V2O3_ARPES_Tdep} shows the detailed evolution of the near-$E_F$ electronic structure, 
when the temperature is first gradually lowered 
from the metallic ($180$~K) to the insulating ($60$~K) state, 
panels (A-F) (curvature of intensity maps) and (L-Q) (raw data),
then increased back to $180$~K, panels (G-K) and (R-V).
In the metallic state at $T \gtrsim 160$K, Figs.~\ref{Fig:V2O3_ARPES_Tdep}(A,~B) and (L,~M), 
the QP band can be described by a free-electron-like parabola
of effective mass $m^{\star} \approx 3.5 m_e$ ($m_e$ is the free-electron mass),
with its band bottom at $E_b \approx -400$~meV.
The quasi-localized state, of weak intensity, can be better seen in the raw data 
around the $\textrm{Z}$ points (near the edges of the energy-momentum maps,
see also Figs.~\ref{Fig:V2O3_ARPES_RawInPlane},~\ref{Fig:V2O3_EDCs_QP_QLS},
~\ref{Fig:V2O3_ARPES_EDCs_July2020},~Supplementary Information),
beyond the Fermi momenta of the QP band. 
Its position, at $E - E_F \gtrsim -240$~meV, is indicated by the red markers in panels (L,~M).
The MH/OV state at $E - E_F \approx -1.1$~eV is also visible --black markers in panels (L,~M).
As the sample is cooled down and enters the transition regime at $130$~K and $120$~K, 
Figs.~\ref{Fig:V2O3_ARPES_Tdep}(C,~D) and (N,~O),
the spectral weight of the QP band decreases,
without any noticeable change in its effective mass (i.e., in its energy-momentum dispersion). 
Simultaneously, the QLS shifts down to $E \lesssim -400$~meV,
becoming more intense as temperature is further lowered.
In the insulating state at $T \lesssim 100$~K, 
Figs.~\ref{Fig:V2O3_ARPES_Tdep}(E,~F) and (P,~Q),
the QP band has vanished. One observes only the weakly but clearly dispersing QLS 
at $E - E_F \approx -700$~meV, 
which now shows a shallow band minimum at $\Gamma$ and maxima
at the monoclinic Brillouin zone edges (see also Fig.~\ref{Fig:V2O3_ARPES_Gral}(F)), 
and the MH/OV band at $E \approx -1.1$~eV. 

Upon heating up from the insulating state at $60$~K back to the metallic state at $180$~K,
Figs.~\ref{Fig:V2O3_ARPES_Tdep}(F-K) and (Q-V), the spectral weight of the QLS decreases, 
rapidly shifting up in energy between $130$~K and $160$~K, 
while the dispersive QP band reappears.
Strikingly, a clear \emph{hysteresis} in the \emph{thermal evolution}
of the electronic states of the system is present, 
best seen in the transition regime around $120$~K and $130$~K 
--compare Figs.~\ref{Fig:V2O3_ARPES_Tdep}(C,~D) to \ref{Fig:V2O3_ARPES_Tdep}(I,~H)
and Figs.~\ref{Fig:V2O3_ARPES_Tdep}(N,~O) to \ref{Fig:V2O3_ARPES_Tdep}(T,~S).
Thus, in the cooling cycle, the QLS becomes more apparent (with respect the QP band)
below $120$~K, temperature at which the QP band is still visible.
while in the heating cycle the QLS is clearly visible until $130$~K,
temperature at which the QP band only starts to re-emerge.
The energy shift of the QLS also shows differences between the cooling and heating cycles,
best seen when comparing the data at $130$~K and $120$~K.
The Supplementary Information presents additional data and analyses
of the thermal evolution of the near-$E_F$ electronic structure measured in different samples.

The observation of a hysteresis in the ARPES spectra is related to the formation 
of phase-domains in the sample, an intrinsic characteristic of first-order phase transitions.
Such domains, of micrometer to sub-micrometer size, have been directly imaged 
in real-space in VO$_2$ and V$_2$O$_3$ by near-field 
infrared microscopy~\cite{Qazilbash-2008,McLeod-2016}, 
photoemission microscopy~\cite{Lupi-2010}
and muon spin-relaxation~\cite{Frandsen-2016}.
As the UV spot used in our experiments has a mean diameter larger than about 30~$\mu$m, 
the ARPES signal is a superposition of electrons emitted from both 
metallic and insulating phases. 
We then approximate the observed ARPES intensity $I(E,k,T)$ at a temperature $T$ 
as a superposition of the intensity measured 
in the pure metallic phase ($T=180$~K) at each energy $E$ and wave-vector $k$
and the intensity measured in the pure insulating phase ($T=60$~K)
at the \emph{same} energy and momentum.
In doing so we are assuming that, as suggested from Fig.~\ref{Fig:V2O3_ARPES_Tdep},
the energy shift of the QLS proceeds rather abruptly with temperature. 
We also neglect thermal broadening, as our energy resolution (of about $15$~meV) 
is already comparable to $k_{B}T$ at $180$~K.
Thus, we write:
\begin{equation}
	\begin{split}
		I(E, k, T) = &a(T) \times I(E, k, T=180{\rm K}) \\
 		&+ b(T) \times I(E, k, T=60{\rm K}) 
	\end{split}
\label{eq:Fit_PES_MIT}
\end{equation}

\begin{figure}[p!]
	\centering
        \includegraphics[clip, width=0.9\linewidth]{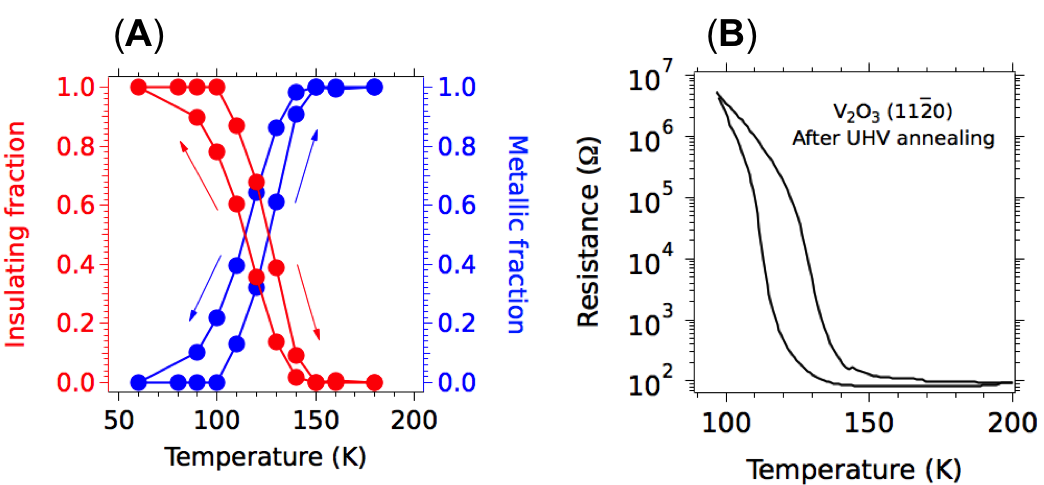}
    \caption{\label{Fig:V2O3_ARPES_hysteresis}
    		 \footnotesize{
    		 \textbf{Fraction of metallic and insulating electronic domains from ARPES.}
			 (A)~Representation of the fraction of insulating and metallic 
			 domains contributing to the ARPES intensity in a
			 V$_2$O$_3$/Al$_2$O$_3$\hkl(11-20) thin film. 
			 Error bars in the fits, Eq.~\ref{eq:Fit_PES_MIT},
			 are smaller than the size of the symbols.
			 (b)~Resistance as a function of temperature measured in the same
			 sample \emph{after} UHV annealing and photoemission experiments.
			 The shift in transition temperature and the decrease of the
			 resistivity in the insulating phase with respect to the 
			 pristine sample, Fig.~\ref{Fig:V2O3_General}(C),
			 are ascribed to the slight doping with oxygen vacancies 
			 induced by annealing in UHV 
			 (Supplementary Information).
        	 }
        }
\end{figure}

Using linear regression, we could therefore determine the phase fractions
$a(T) \geq 0$ and $b(T) \geq 0$  
that best fit the measured spectra \emph{for all energy and momenta},
i.e., over a set of around $4 \times 10^{5}$ points of $I(k,E)$, at each temperature.
Fig.~\ref{Fig:V2O3_ARPES_hysteresis}(A) shows the so-calculated
fraction of insulating and metallic domains, $b/(a+b)$ and $a/(a+b)$, 
as a function of temperature.
For comparison, the resistance obtained on the same sample \emph{after} ARPES measurements,
Fig.~\ref{Fig:V2O3_ARPES_hysteresis}(B), is also shown.
The agreement in the hysteresis between the electrical resistance 
and the ARPES data (onset of the transition on cooling at around $140-150$~K,
mid-point at around $120$~K, thermal amplitude of the hysteresis 
of about $15-20$~K)
indicates that the observed changes in the spectral function
are directly linked to the metal-to-insulator transition.

Our temperature-dependent studies were performed along $k_{<111>}$, 
a direction orthogonal to the antiferromagnetic wave-vector 
in the insulating state~\cite{Moon-1970,Bao-1997}.
Hence the band dispersions along this direction should not be directly affected 
by antiferromagnetic band folding. 
In line with this expectation, our ARPES data do not show any folding 
of the dispersive QP state. Instead, the spectral weight of the QP state vanishes 
as the system goes from metallic to insulating, without any measurable change 
in its dispersion, while the quasi-localized state shifts down in energy 
and increases in spectral weight. 
One possibility is that the energy shift of the QLS, 
which remains essentially non-dispersing along all high-symmetry directions 
explored in this work (Fig.~\ref{Fig:V2O3_ARPES_Tdep} and 
Supplementary Information, Fig.~\ref{Fig:V2O3_ARPES_Spaguetthi}), 
is associated to its nesting along momenta parallel to the antiferromagnetic wave-vector. 
On the other hand, note that in the insulating state the magnetic moments 
are ordered ferromagnetically along $k_{<111>}$~\cite{Moon-1970}.
This of course can affect the band structure, although in a way different 
from antiferromagnetic folding. 
Future theoretical studies should address further how the specific magnetic ordering 
of V$_2$O$_3$ affects its different orbital states.

More generally, our ARPES measurements show that, across the MIT, 
the essential redistribution of spectral weight in the occupied part 
of the electronic spectrum occurs between the QP and QLS bands, 
over an energy range of about $700$~meV below $E_F$. 
On the other hand, recent optical measurements show that, 
while a large suppression of optical conductivity occurs indeed 
over an energy range of about $1$~eV~\cite{Qazilbash-PRB-2008,Stewart-2012,Lo-Vecchio-2015},
a significant spectral weight transfer extends up to and beyond 
at least $3$~eV~\cite{Qazilbash-2008,Stewart-2012}.
Thus, taken together, ARPES and optics data indicate that the unoccupied part 
of the electronic spectrum is also undergoing a major reconstruction,
over a large energy range of several eV, across the MIT. 

Our observations of spectral weight redistribution among the QP and QLS
are consistent with X-ray absorption spectroscopy measurements,
which found that the metallic and insulating phases
have different orbital occupancies among two states
present in both phases~\cite{Park-2000,Rodolakis-2010}.
Our findings are also in line with multi-orbital 
first-principles calculations for V$_2$O$_3$~\cite{Poteryaev-2007,Grieger-2015,Lechermann-2018}.
In the metallic phase, the dispersive quasi-particle band 
seen in the ARPES spectra and crossing the Fermi level 
is identified in the calculations as a band of $a_{1g}$ dominant character, 
while the remaining spectral weight stems essentially from the $e_g^{\pi}$ orbitals. 
In particular, the latter would be mainly responsible for the spectral weight 
of the broad quasi-localized state around $E_F$,  
as well as the Hubbard band feature around $-1.1$~eV.
In the insulating phase, the dispersive $a_{1g}$ band would be emptied, and only
the double peak structure (at energies $-0.7$~eV and $-1.1$~eV) 
of the Hubbard band formed by the now half-filled $e_g^{\pi}$-states 
would survive~\cite{Poteryaev-2007,Poteryaev-2008,Grieger-2015}.
Our data provides thus a direct, momentum-resolved illustration of the multi-orbital nature
of the Mott transition in V$_2$O$_3$.



\newpage

%
\newpage
%
%
\newpage
%
%
\newpage
%

\clearpage

\appendix

\section*{METHODS}
\label{Secion:Materials and Methods}
\subsection{Thin-film growth and characterization}
The 100~nm thick epitaxial V$_2$O$_3$ thin films were deposited on 
\hkl(11-20) or \hkl(01-12) oriented Al$_2$O$_3$ substrates by RF magnetron sputtering, 
as described in a previous work~\cite{Kalcheim-2019}.
An approximately 8~mTorr ultra high purity Argon ($>99.999$\%) 
and the growth temperature of 700$^{\circ}$C 
were used during the sputtering process. The thin film structural properties 
were measured using a Rigaku SmartLab x-ray diffraction system. 
Resistance vs temperature measurements were done using a TTPX Lakeshore cryogenic probe station 
equipped with a Keithley 6221 current source and a Keithley 2182A nanovoltmeter.

\subsection{Crystal structure and Brillouin zone of V$_2$O$_3$}
\label{struc_BZ}
In the high-temperature metallic phase, bulk V$_2$O$_3$ crystallizes 
in a rhombohedral (corundum) structure, as shown in Fig.~\ref{Fig:V2O3_General}(B),
with lattice parameters $|\mathbf{a}|=|\mathbf{b}|=|\mathbf{c}|=5.467$~\AA~and
angles between lattice vectors $\alpha=\beta=\gamma=53.74^{\circ}$.
Such a structure can be equally represented in a non-primitive hexagonal cell,
also shown in Figs.~\ref{Fig:V2O3_General}(B,~C).
The Miller indices $(h_R, k_R, l_R)$ in rhombohedral coordinates
are related to the Miller indices $(h_H, k_H, i_H=-h_H-k_H, l_H)$ 
in the corresponding hexagonal coordinates by:
\begin{eqnarray*}
	h_R & = & \frac{1}{3} (-k_H+i_H+l_H) \\
	k_R & = & \frac{1}{3} (h_H-i_H+l_H) \\
	l_R & = & \frac{1}{3} (-h_H+k_H+l_H) \\ 
\label{eq:Miller_indices}	
\end{eqnarray*}

Our thin films are grown on Al$_2$O$_3$\hkl(11-20) (a-plane) 
and Al$_2$O$_3$\hkl(01-12) (R-plane) surfaces,
corresponding respectively to the \hkl(-110) and \hkl(011) planes
in rhombohedral coordinates.
Thin films grown on different Al$_2$O$_3$ surfaces present a slight distortion
(at most 1\% of strain) from the ideal corundum symmetry~\cite{Kalcheim-2019}, 
which is negligible for ARPES measurements. Hence, in this work,
we use the bulk primitive rhombohedral Brillouin zone to represent the ARPES data.

\subsection{ARPES measurements}
ARPES experiments were performed at the CASSIOPEE beamline of Synchrotron SOLEIL (France) 
and at beamline 2A of KEK-Photon Factory (KEK-PF, Japan) 
using hemispherical electron analyzers with vertical and horizontal slits, respectively. 
Typical electron energy and angular resolutions were $15$~meV and 0.25$^{\circ}$. 
In order to generate pristine surfaces for the ARPES experiments, 
the thin films of V$_2$O$_3$ were annealed for 5-10 minutes 
at approximately 550-600$^{\circ}$C in UHV conditions,
at a base pressure of $10^{-9}$~mbar before annealing, 
reaching a desorption peak of $\approx 8 \times 10^{-7}$~mbar during annealing.
Low-energy electron diffraction (LEED) was employed to verify the long-range crystallinity 
and cleanliness of our surfaces after preparation 
--see Supplementary Information, Fig.~\ref{Fig:V2O3_LEED}. 

While the insulating phase of V$_2$O$_3$ would prevent the realization 
of ARPES experiments on a bulk sample, 
we found that the residual conductivity in our samples 
was enough to reduce the charging of the film to approximately 
$11$~eV at $100$K under the photon flux used for our measurements. 
The charging was subsequently corrected by aligning the energy of 
the Fermi-Dirac steps (if a residual spectral weight at $E_F$ was present)
or of the valence bands at all temperatures, thus assuming that such a band
is not affected by the MIT. Such a hypothesis is not only physically sound,
given the high binding energy of the O-$2p$ levels that form the VB and their 
large separation from the MH and QP bands, but is also supported by the data which, 
apart from the expected thermal broadening and energy shift due to charging, 
do not show any significant changes in the valence band shape
when the system is cooled/warmed across the MIT.

\subsection{Second derivative rendering of ARPES intensity maps}
For second derivative rendering 
(Supplementary Information, Fig.~\ref{Fig:V2O3_ARPES_FermiSurfaces}(C))
the raw photoemission energy-momentum maps were convolved
with a two-dimensional Gaussian of FWHM's $2^{\circ}$ in angle and $25$~meV in energy
in order to smooth the noise.  
Only negative values of the second derivative, 
representing peak maxima in the original data, are shown.

\subsection{2D curvature rendering of ARPES intensity maps}
The 2D curvature method~\cite{Zhang2011} was used to enhance the intensity of broad/weak 
spectral features in the ARPES intensity maps. 
To this end, boxcar smoothing was applied to the raw data.
For Figs.~\ref{Fig:V2O3_ARPES_Gral}(B,~C), 
we used a kernel of 80~meV~$\times 0.037$~\AA$^{-1}$
with a 2D curvature free parameter of $0.2$.
For Figs.~\ref{Fig:V2O3_ARPES_Tdep}(A-K), 
we used a kernel of 140~meV~$\times 0.026$~\AA$^{-1}$
with a 2D curvature free parameter of $0.1$.
Only negative values of the 2D curvature, which represent maxima in the original data, 
are shown.
After taking the 2D curvature, the data 
were rigidly shifted by 30~meV towards the Fermi level, 
in order to preserve the exact peak positions as determined by the raw data 
and its second derivatives. 
There is no effect on the conclusions of this study as an identical shift was applied 
to all figures where the 2D curvature rendering is displayed.

\subsection{3D $k$-space mapping}
Within the free-electron final state model,
ARPES measurements at constant photon energy give the electronic structure at the surface of 
a spherical cap of radius 
$k = \sqrt{\frac{2m_\text{e}}{\hbar^2}} \left(h\nu - \Phi + V_0 \right)^{1/2}$. 
Here, $m_\text{e}$ is the free electron mass, $\Phi$ is the work function, 
and $V_0 = 12.5$~eV is the ``inner potential" of V$_2$O$_3$~\cite{Lo-Vecchio-2016}.
Measurements around normal emission provide the electronic structure in a plane 
nearly parallel to the surface plane.
Likewise, measurements as a function of photon energy provide the electronic structure in a plane
perpendicular to the surface.

\subsection{Thermal cycling in ARPES measurements}
In order to accurately capture the electronic hysteresis behavior across the MIT,
the samples were loaded into the ARPES manipulator at $200$K. Once thermalized,
we performed a slow stepwise cooling cycle,
setting the parameters in the temperature controller so as to avoid overshooting
the desired new temperature (hence avoiding spurious hysteresis cycles), 
then letting the system thoroughly thermalize at that temperature for over 30~minutes
before measuring. After reaching the lowest measurement temperature, 
we performed an analogous heating cycle, measuring at the same temperatures
as during the cooling cycle.

\subsection{Acknowledgments} 
We thank Fran\c cois~Bertran for assistance during ARPES
measurements at CASSIOPEE (Synchrotron SOLEIL).
Work at ISMO was supported by public grants from the French National Research Agency (ANR), 
project Fermi-NESt No. ANR-16-CE92-0018, 
the ``Laboratoire d'Excellence Physique Atomes Lumi\`ere Mati\`ere'' 
(LabEx PALM projects ELECTROX, 2DEG2USE and 2DTROX) overseen by the ANR as part of the 
``Investissements d'Avenir'' program (reference: ANR-10-LABX-0039),
and the CNRS International Research Project EXCELSIOR.
Work at KEK-PF was supported by Grants-in-Aid for Scientific Research 
(Nos. 16H02115 and 16KK0107) from the Japan Society for the Promotion of Science (JSPS).
Experiments at KEK-PF were performed under the approval of the 
Program Advisory Committee (Proposals 2016G621 and 2018S2-004) 
at the Institute of Materials Structure Science at KEK.
Work at UCSD was supported by the Air Force Office of Scientific Research
under award number FA9550-20-1-0242.
R.~L.~B. was supported by NWO through a VICI grant.
P.~H.~R.-G. was supported by a CAPES grant, Brazil.

\subsection{Author contributions}
Project conception: A.F.S.-S., M.J.R., J.T. and I.K.S.;
ARPES measurements: M.T, R.L.B., P.H.R.-G., E.D., E.F., F.F., P.L.F., K.H., 
H.K, and A.F.S.-S;
sample growth, structural characterization and transport measurements: M.-H.L., N.M.V., Y.K., 
under the supervision of I.K.S;
infrared measurements: A.Z.; 
data analysis and interpretation: M.T, E.F and A.F.S.-S, with input from M.J.R. and S.B.;  
writing of the manuscript: M.T. and A.F.S.-S.
This is a highly collaborative research. 
All authors discussed extensively the results, interpretation and manuscript.

\subsection{Competing interests}  
Authors declare that they have no competing financial interests.

\subsection{Data availability}
All data needed to evaluate the conclusions of this work 
are present in the paper and/or the Supplementary Information. 

\subsection{Additional information}
Correspondence and requests for materials should be addressed to 
A.F.S.-S. (e-mail:~andres.santander-syro@universite-paris-saclay.fr).


\clearpage

\renewcommand{\thefigure}{S\arabic{figure}} 
\setcounter{figure}{0}

\section*{SUPPLEMENTARY INFORMATION}

\subsection{Low-energy electron diffraction (LEED)}
\begin{figure}[t!h]
	\centering
        \includegraphics[clip, width=0.9\linewidth]{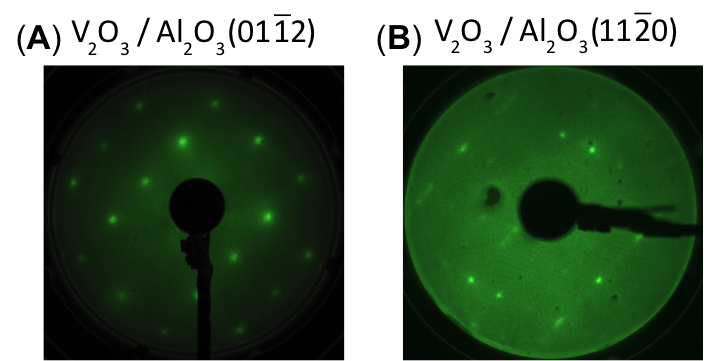}
    \caption{\footnotesize{
        	 \textbf{LEED images of the studied V$_2$O$_3$ thin film surfaces.}
        	 (A,~B)~LEED images of V$_2$O$_3$/Al$_2$O$_3$\hkl(01-12)
        	 and V$_2$O$_3$/Al$_2$O$_3$\hkl(11-20) films, respectively, 
        	 obtained after in-situ annealing, right before ARPES measurements.
        	 }
        }
\label{Fig:V2O3_LEED}
\end{figure}
Figure~\ref{Fig:V2O3_LEED} shows the LEED images of our V$_2$O$_3$/Al$_2$O$_3$\hkl(01-12)
and V$_2$O$_3$/Al$_2$O$_3$\hkl(11-20) thin films obtained after in-situ annealing. 
The periodically arranged sharp spots indicate a clean, crystalline surface, 
suitable for ARPES measurements.

\subsection{Effects of in-situ annealing on the V$_2$O$_3$ thin films}
\label{effect_anneal}
\begin{figure}[t!h]
	\centering
        \includegraphics[clip, width=0.9\linewidth]{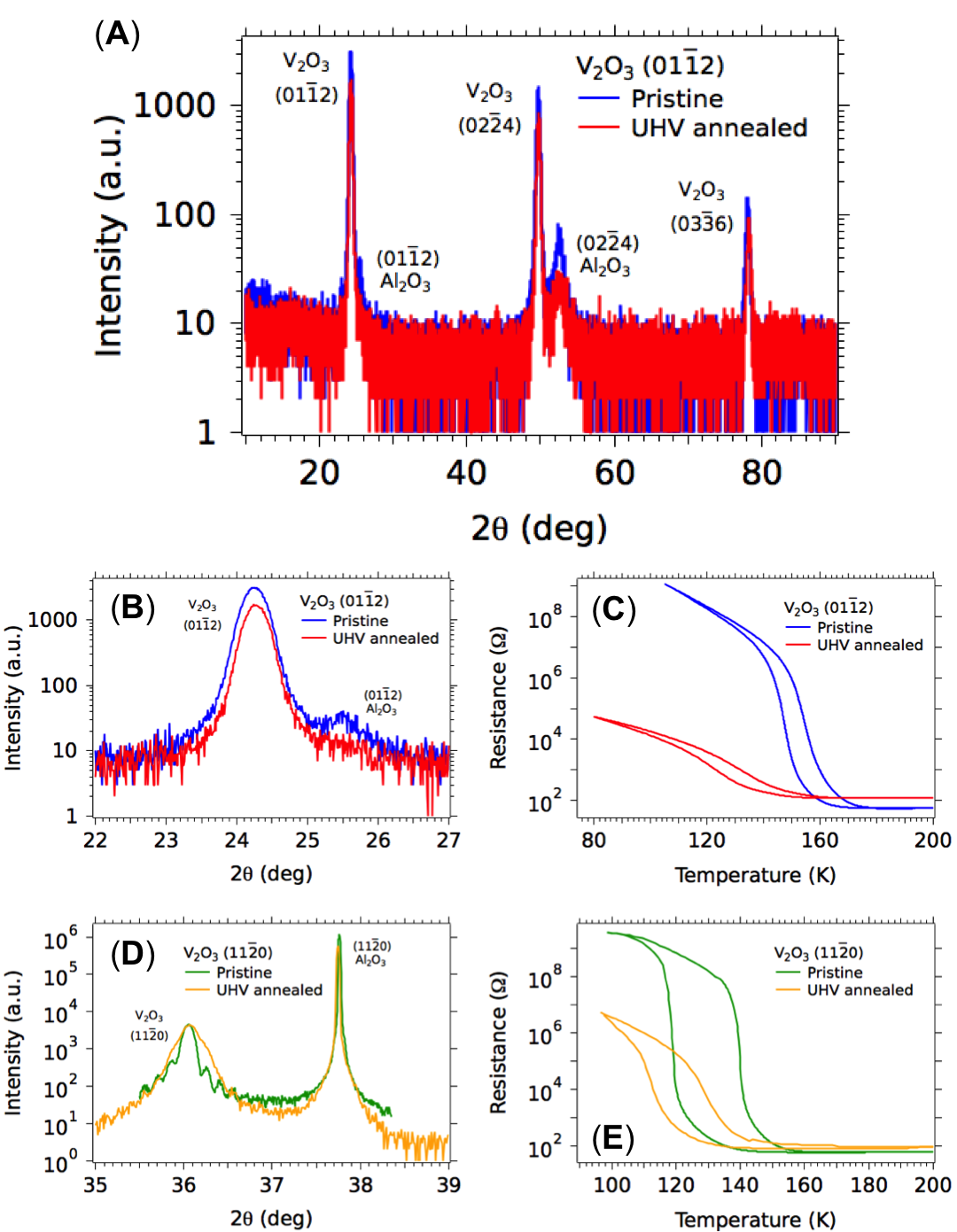}
    \caption{\footnotesize{
        	 \textbf{X-ray and resistance characterizations of the V$_2$O$_3$ films.}
        	 (A,~B)~X-ray diffraction characterization, and (C)~resistance measurements
        	 on a V$_2$O$_3$/Al$_2$O$_3$\hkl(01-12) thin film 
        	 used in some of our ARPES experiments
        	 (see Fig.~\ref{Fig:V2O3_PES_hysteresis} below),
        	 before (blue) and after (red) the in-situ annealing in UHV. 
        	 Panel (B) is a zoom over the V$_2$O$_3$ and Al$_2$O$_3$ \hkl(01-12) peaks.
			 (D,~E)~Analogous measurements to (B,~C) 
			 on the V$_2$O$_3$/Al$_2$O$_3$\hkl(11-20) thin films 
        	 used in our ARPES experiments, 
        	 before (green) and after (yellow) the in-situ annealing in UHV. 
        	 }
        }
\label{Fig:V2O3_XRD_RT_Annealing}
\end{figure}

Fig.~\ref{Fig:V2O3_XRD_RT_Annealing} presents the X-ray diffraction (XRD) and 
electrical resistance characterizations of our V$_2$O$_3$/Al$_2$O$_3$\hkl(01-12) and 
V$_2$O$_3$/Al$_2$O$_3$\hkl(11-20) thin films before (blue or green curves) 
and after (red or yellow curves) the annealing in UHV required for ARPES experiments.
Note that XRD is a bulk-sensitive measurement probing the entire thickness 
of the V$_2$O$_3$ film, while temperature-dependent electrical transport properties 
are very sensitive to changes in surface oxygen stoichiometry. 
Accordingly, no significant changes in crystal structure 
(main peaks of the XRD diagrams), hence in the bulk stoichiometry of the film, are observed.
On the other hand, the onset of the resistive MIT on cooling is shifted down in temperature 
from $\approx 160$K to $\approx 140$~K in the V$_2$O$_3$/Al$_2$O$_3$\hkl(01-12) film,
and from  $\approx 140$~K to $\approx 130$~K in the V$_2$O$_3$/Al$_2$O$_3$\hkl(11-20) film,
while the change in resistance between the insulating and metallic phases
is reduced by about three orders of magnitude in both films.

We ascribe the observed changes after UHV annealing to the creation of oxygen vacancies,
a phenomenon well known and reported in many other transition metal 
oxides~\cite{Roedel-2016, Roedel-2017, Loemker-2017, Roedel-2018, Santander-2011, Santander-2012},
including other correlated-electron vanadium oxides such as SrVO$_3$~\cite{Backes-2016}.
Indeed, in many of these oxides oxygen vacancies act as electron 
donors~\cite{Roedel-2016,Backes-2016,Roedel-2017,Loemker-2017,Roedel-2018,Santander-2011,
Santander-2012}.
The important fact, however, is that the MIT and its hysteresis cycle are preserved, 
as indeed attested by the ARPES data, the latter being in very good agreement
with the resistivity hysteresis cycle after UHV annealing.
Such slight metallization of the sample is beneficial for the ARPES measurements,
as it allows following the MIT without excessive charging down to the insulating state.

\begin{figure}[t!h]
	\centering
        \includegraphics[clip, width=0.9\linewidth]{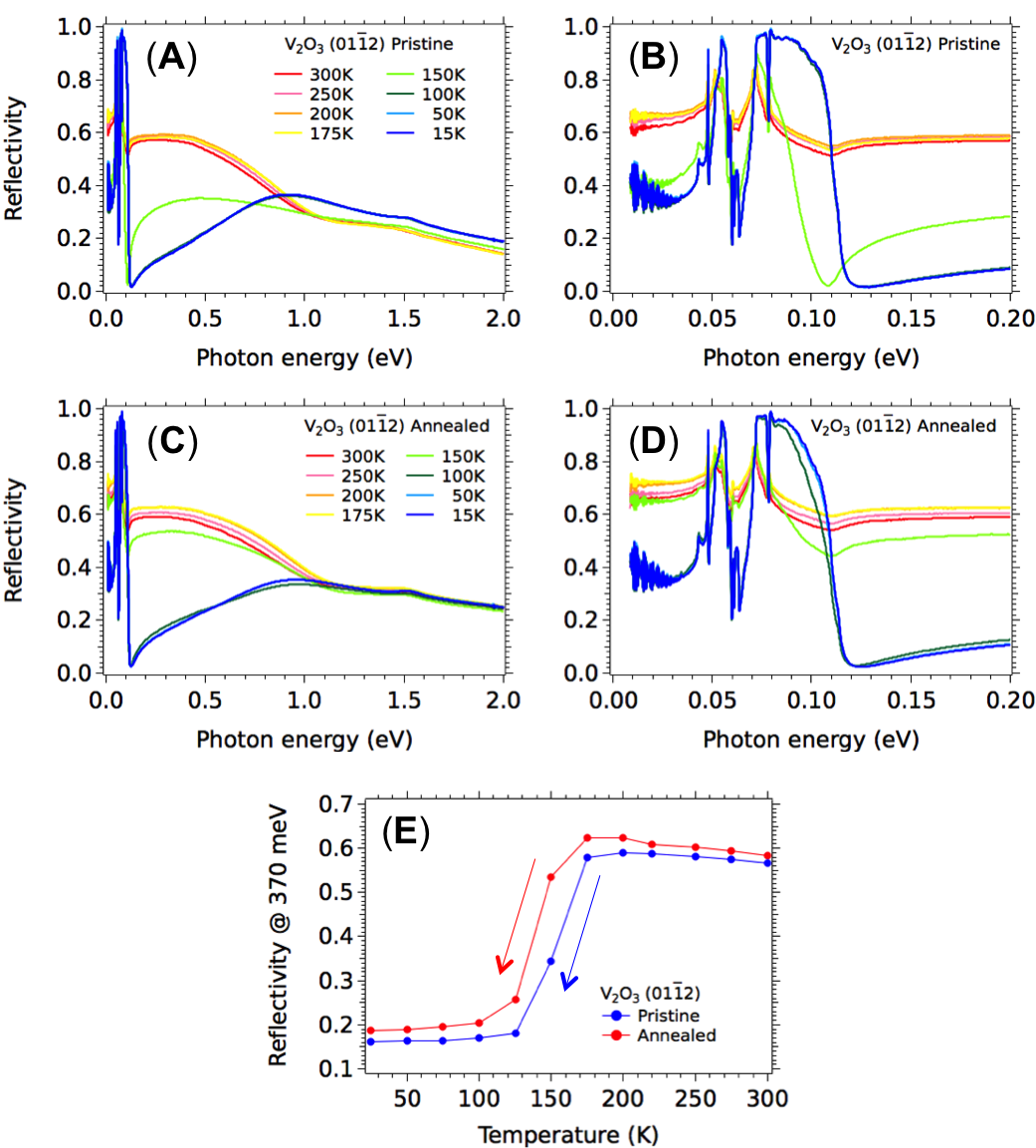}
    \caption{\footnotesize{
        	 \textbf{Infrared reflectivity characterization of the V$_2$O$_3$ films.}
        	 (A)~Infrared reflectivity measurements for various temperatures
        	 (cooling cycle) in a pristine V$_2$O$_3$/Al$_2$O$_3$\hkl(01-12) 
        	 thin film.
        	 (B)~Zoom of the previous data over the low-energy excitations.
			 (C,~D)~Similar to (A,~B) on a sample of the same batch that was
			 annealed in UHV prior to ARPES measurements.
			 (E)~Reflectivity at $370$~meV as a function of temperature
			 for both samples. 
			 All these data show that the MIT is still present in the annealed sample 
			 studied by ARPES, and the onset of the transition
			 has just shifted down in temperature by about $15$~K.
        	 }
        }
\label{Fig:V2O3_IR_Annealing}
\end{figure}

The previous is confirmed by infrared spectroscopy measurements, 
Fig.~\ref{Fig:V2O3_IR_Annealing}. These show that, in the metallic phase, 
the low-energy (far infrared) reflectivity, 
related to the concentration of free carriers in the sample, 
is slightly lower in a pristine V$_2$O$_3$/Al$_2$O$_3$\hkl(01-12) sample 
compared to a sample of the same batch that was annealed then measured by ARPES. 
Note that, in the insulating phase, the reflectivity of both samples 
strongly decreases at energies below about $800-900$~meV, 
showing thus that the optical gap in our thin films is similar 
to the one measured in single crystals, and is not appreciably affected by the UHV annealing. 
The only appreciable effect of UHV annealing, 
in phase with our characterizations from resistivity measurements, 
is that the temperature-dependent change in reflectivity at, e.g., $370$~meV 
(inside the energy gap), is shifted down by about $15$~K in the annealed sample.

\subsection{Rhombohedral and monoclinic Brillouin zones of V$_2$O$_3$}
\label{Comparison_BZs}
\begin{figure}[t!h]
	\centering
        \includegraphics[clip, width=0.9\linewidth]{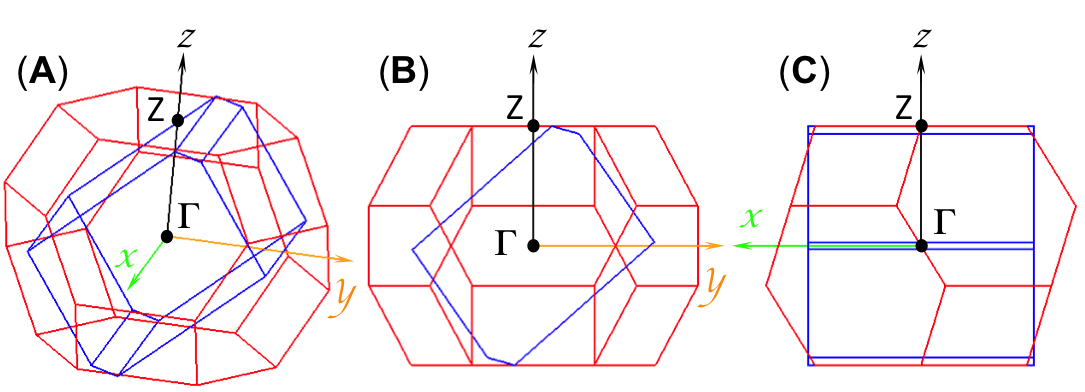}
    \caption{\footnotesize{
        	 \textbf{Rhombohedral and monoclinic Brillouin zones of V$_2$O$_3$.}
        	 (A)~Comparison between the rhombohedral (red) and monoclinic (blue) 
        	 Brillouin zones of V$_2$O$_3$, corresponding respectively to the 
        	 metallic and insulating phases. The $xyz$ axes are also shown. 
        	 (B,~C)~Same as (A) viewed from the $x$-side and $y$-side, 
        	 respectively.
        	 }
        }
\label{Fig:V2O3_Comparison_BZs}
\end{figure}

Figure~\ref{Fig:V2O3_Comparison_BZs} compares 
the rhombohedral and monoclinic Brillouin zones 
in the metallic and insulating phases of V$_2$O$_3$, respectively. 
While the two Brillouin zones are quite different in 3D reciprocal space, 
our temperature-dependent data were all measured in a fixed direction 
(in the laboratory frame of reference) coinciding with the rhombohedral 
$\Gamma$Z direction. Along this direction, the distance from $\Gamma$ 
to the Brillouin zone edge is reduced only by about $0.09$~\AA$^{-1}$~in the 
monoclinic phase (i.e., a reduction of about 13\% with respect the 
rhombohedral $\Gamma$Z  distance). 
This value is comparable to our experimental resolution in momentum. 
Thus, for simplicity in notation, we choose to refer all momentum directions 
in our ARPES data to the rhombohedral ``frame of reference''. 
When pertinent in a given figure, the monoclinic zone edges 
for the energy-momentum maps in the insulating state are also shown.

\subsection{In-plane and out-of-plane Fermi surfaces}
\label{ARPES_FS_InPlane_OutOfPlane}
\begin{figure*}[t!h]
	\centering
        \includegraphics[clip, width=0.8\linewidth]{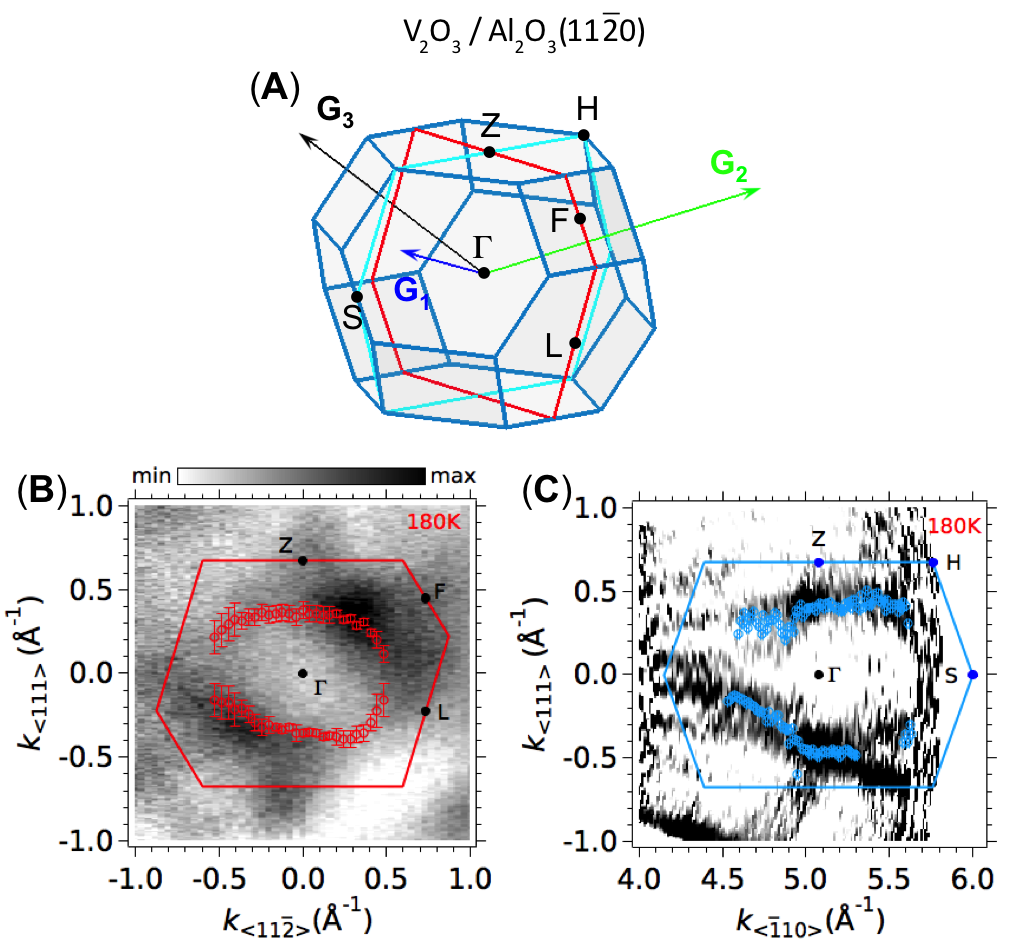}
    \caption{\footnotesize{
    		 \textbf{In-plane and out-of-plane Fermi surfaces in the metallic state of V$_2$O$_3$.}
        	 (A)~Rhombohedral 3D Brillouin zone of V$_2$O$_3$, 
        	 showing the primitive vectors of the
        	 reciprocal lattice, together with a \hkl(-110) plane (red) 
        	 and a \hkl(11-2) plane (light blue). These planes are, respectively,
        	 parallel and perpendicular to the surface of the 
        	 studied V$_2$O$_3$/Al$_2$O$_3$\hkl(11-20) films.
        	 (B)~Same in-plane Fermi-surface map (metallic state at $180$~K)
        	 of a V$_2$O$_3$/Al$_2$O$_3$\hkl(11-20) sample shown in the main text. 	 
        	 Red open markers show the experimental Fermi momenta with
        	 their respective error bars. Data were measured at $h\nu = 86$~eV
			 (c)~Out-of-plane Fermi-surface map (2D curvature, see Methods)
        	 of the same sample. Blue open markers show the experimental Fermi momenta with
        	 their respective error bars. 
        	 Data were acquired by varying the photon energy between $55$~eV and $120$eV 
        	 in steps of $0.5$~eV using linear horizontal polarized light.
        	 }
        }
\label{Fig:V2O3_ARPES_FermiSurfaces}
\end{figure*}

Figure~\ref{Fig:V2O3_ARPES_FermiSurfaces} shows the in-plane and out-of-plane Fermi surfaces
measured at $180$~K for the same V$_2$O$_3$/Al$_2$O$_3$\hkl(11-20) studied in the main text,
together with the experimentally determined Fermi momenta. The out-of-plane dispersion
in the Fermi surface corroborates the 3D nature of the measured state,
as 2D surface states would not disperse in the momentum direction 
perpendicular to the surface. 
The main feature of the observed Fermi surface is a large electron-like sheet 
centered at $\Gamma$ -however open in the direction $\Gamma \textrm{L}$, as will be seen next.

\subsection{Raw ARPES energy-momentum maps}
\label{ARPES_Raw_Ek}
\begin{figure*}[t!h]
	\centering
        \includegraphics[clip, width=0.8\linewidth]{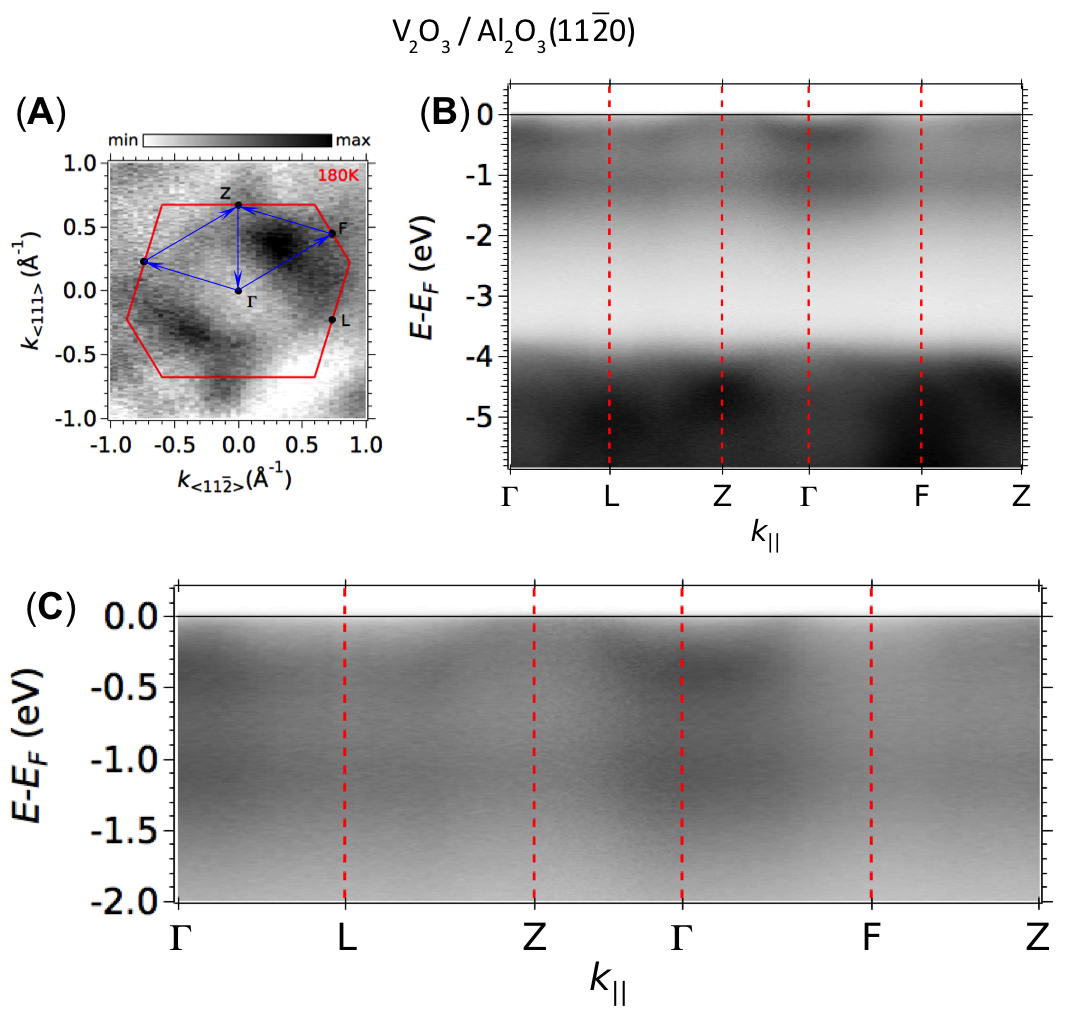}
    \caption{\footnotesize{
    		 \textbf{Raw ARPES energy-momentum maps in the metallic state of V$_2$O$_3$.}
        	 (A)~Same in-plane Fermi-surface map as shown in previous figures
        	 indicating the path in reciprocal space for the raw ARPES energy-momentum
        	 maps in the metallic state ($180$~K) presented in panels (B,~C). 
        	 Data in (B) include the VB-top, MH/OV, QP and QLS bands. 
        	 Panel (C) is a zoom of (B) over the MH/OV, QP and QLS features.
        	 }
        }
\label{Fig:V2O3_ARPES_Spaguetthi}
\end{figure*}

Figure~\ref{Fig:V2O3_ARPES_Spaguetthi}(A) shows the 
$\Gamma-\textrm{L}-\textrm{Z}-\Gamma-\textrm{F}-\textrm{Z}$ path in reciprocal space,
and Figs.~\ref{Fig:V2O3_ARPES_Spaguetthi}(B,~C) show the ARPES energy-momentum maps 
in the metallic phase along such a path. 
One observes that the QP state has an electron-like dispersion along $\Gamma \textrm{Z}$
and $\Gamma \textrm{F}$, while along $\Gamma \textrm{L}$ the state appears broad and,
within resolution, non dispersive -such that this Fermi surface sheet would be open around
the $\textrm{L}$ points. The state shows a hole-like dispersion along $\textrm{L} \textrm{Z}$,
such that around $\textrm{Z}$ one has a hole-like Fermi pocket.
The QLS can be observed as a weak, broad, non-dispersive part of spectral weight near $E_F$
at all momenta, and is more visible around $\textrm{Z}$ and along $\textrm{ZF}\textrm{Z}$. 
Additional data for the QLS, and its thermal evolution, will be discussed in the next section.

\begin{figure*}[t!h]
	\centering
        \includegraphics[clip, width=0.9\linewidth]{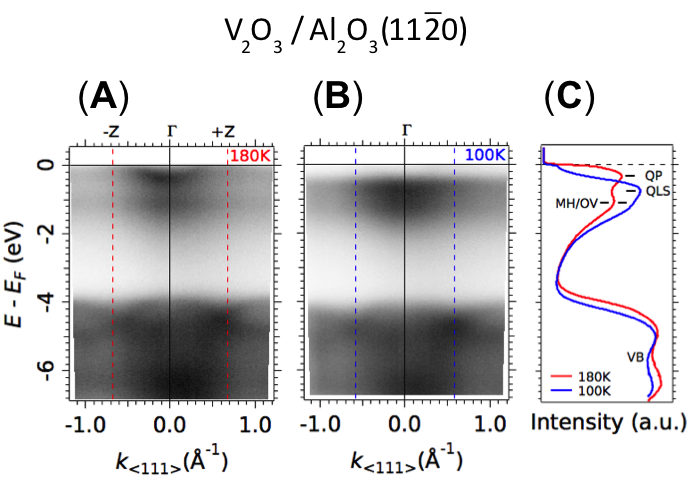}
    \caption{\footnotesize{
    		 \textbf{Raw ARPES energy-momentum maps in the metallic 
    		 and insulating states of V$_2$O$_3$.}
        	 (A,~B)~Raw ARPES energy-momentum maps along $\Gamma \textrm{Z}$ (or $k_{<111>}$) 
        	 over a wide energy range in the metallic (180~K) and insulating (100~K) states 
        	 respectively, corresponding to the 2D curvature images 
			 shown in Figs.~2(B,~C) of the main text.
			 The rhombohedral Brillouin zone edges ($\pm \textrm{Z}$ points) 
			 at $180$~K, and the monoclinic zone edges at $100$~K, 
			 are indicated by red and blue dashed lines, respectively.  
			 (C)~Corresponding momentum-integrated ARPES intensities
			 showing the QP, QLS, MH/OV and VB states, 
			 identical to Figs.~2(D) of the main text,
			 displayed again here for completeness.
			 All data in this figure were measured at a photon energy of $86$~eV,
			 corresponding to a bulk $\Gamma$ point in the out-of-plane direction,
			 using linear horizontal light polarization.
        	 }
        }
\label{Fig:V2O3_ARPES_RawInPlane}
\end{figure*}

Figures~\ref{Fig:V2O3_ARPES_RawInPlane}(A-C) show, respectively, 
the raw energy-momentum ARPES maps along $\Gamma \textrm{Z}$, or $k_{<111>}$,
at 180~K and 100~K, and the integrated ARPES intensity of the previous two, 
corresponding to the same samples and data presented in Fig.~2 of the main text.
At $180$~K, the QP, MH/OV and VB are clearly visible. The weak non-dispersive QLS 
is also seen right below $E_F$ in Fig.~\ref{Fig:V2O3_ARPES_RawInPlane}(A), 
spanning all the momenta within the measurement window.
At $100$~K, the QP band has vanished, and only the QLS, MH/OV, and VB are observed.

\subsection{Detailed temperature evolution of the quasi-localized state}
\label{ARPES_QLS_Temperature}
\begin{figure*}[h!t]
	\centering
        \includegraphics[clip, width=0.8\linewidth]{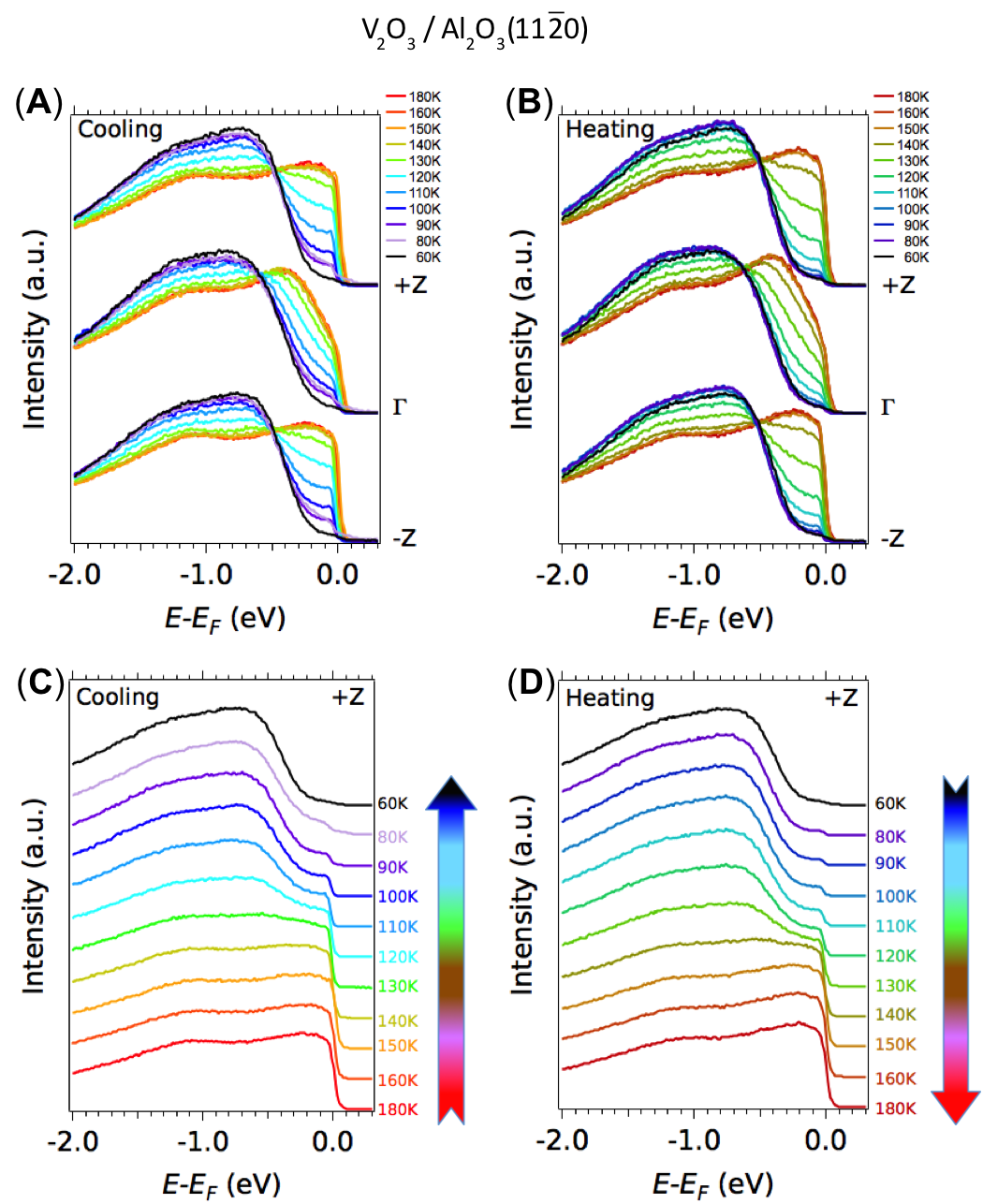}
    \caption{\footnotesize{
        	 \textbf{Temperature evolution of the spectra at $\Gamma$ and $\pm \textrm{Z}$.}
        	 (A,~B)~Energy distribution curves (EDCs) at $\Gamma$ and $\pm \textrm{Z}$ 
        	 (integrated over $0.03$~\AA$^{-1}$ around each point) over, respectively, 
        	 a cooling and a heating cycle across the MIT. 
        	 (C,~dD~Thermal evolution (cooling and heating) of the EDCs
        	 only at the $+\textrm{Z}$ point.
        	 All data in this figure correspond to the spectra shown in Fig.~3 of the main text,
        	 including additional temperatures, for a V$_2$O$_3$\hkl(11-20) thin film. 
        	 Spectra at the same temperature on opposite thermal cycles are shown 
        	 in slightly different color hues, for clarity.
        	 }
        }
\label{Fig:V2O3_EDCs_QP_QLS}
\end{figure*}

In the metallic phase at high temperatures, the quasi-localized state is best observed at momenta 
where it does not overlap with the intense dispersing QP state, such as the edges
of the Brillouin zone --where the QP peak is located at energies above $E_F$.
Figures~\ref{Fig:V2O3_EDCs_QP_QLS}(A,~B) compare the energy distribution curves (EDCs)
at the $\Gamma$ (zone center) and $\pm \textrm{Z}$ (zone edges) points 
over, respectively, a cooling and heating cycle across the MIT.
Figures~\ref{Fig:V2O3_EDCs_QP_QLS}(C,~D) show the temperature
evolution (cooling and heating) of the EDCs only at the $+\textrm{Z}$ point.
The data correspond to the spectra shown in Fig.~3 of the main text,
including additional temperatures.
At $180$~K, the broad peak of the QLS around $E-E_F \approx -240$~meV 
can be clearly observed at the zone edges. 
Note that thermal plus resolution broadening of the Fermi-Dirac cutoff 
occur on a smaller energy scale, of about $20$~meV around $E_F$,
so the peak at $E-E_F \approx -240$~meV corresponds to an intrinsic 
spectral feature.
As temperature decreases, 
the peak remains at about the same position down to $140$~K,
then starts to rapidly shift down in energy until reaching an energy of
$E-E_F \approx -700$~meV at $100$~K, temperature below which the peak
does not evolve further.
Upon heating, the QLS peak at the zone edges stays at $E-E_F \approx -700$~meV 
until about $120$~K. Above this temperature 
the peak shifts rapidly up in energy until reaching its original
high-temperature position at around $150$~K.
 
On the other hand, the EDCs at $\Gamma$ show an intense peak at 
$E-E_F \approx -400$~meV at $180$~K. Its asymmetric line-shape,
broader at energies above the peak towards $E_F$,
corresponds to the superposition of the dispersive QP peak and the QLS.
As temperature decreases and goes below $140$~K, the spectral weight of this peak
starts to rapidly decrease at energies above the peak, while the peak itself
appears to shift down in energy until reaching $E-E_F \approx -800$~meV at $100$~K,
below which the peak does not change appreciably.
From the momentum-resolved data, Fig.~3 of the main text, we know that the 
band minimum and effective mass of the QP state do not change with temperature. 
Only its intensity decreases upon entering the insulating state.
Thus, the thermal evolution of the EDC at $\Gamma$ actually corresponds to
the superposition of two effects: the weight of the QP peak (fixed in energy) 
becomes weaker upon cooling into the insulating phase, while the peak of the QLS
simultaneously shifts down in energy until reaching an energy lower than, 
but close to, the bottom of the original QP state.
Upon heating, the behavior of the EDCs at $\Gamma$ is reversed, however showing a hysteresis:
only at temperatures above $120$~K the peak starts to shift up in energy 
and gain spectral weight at energies higher that its position, until reaching 
its final energy position and lineshape at or above $150$~K.

The energy of the MH/OV state at $E-E_F \approx -1.1$~eV remains temperature-independent
within our resolution.

\begin{figure*}[h!t]
	\centering
        \includegraphics[clip, width=0.9\linewidth]{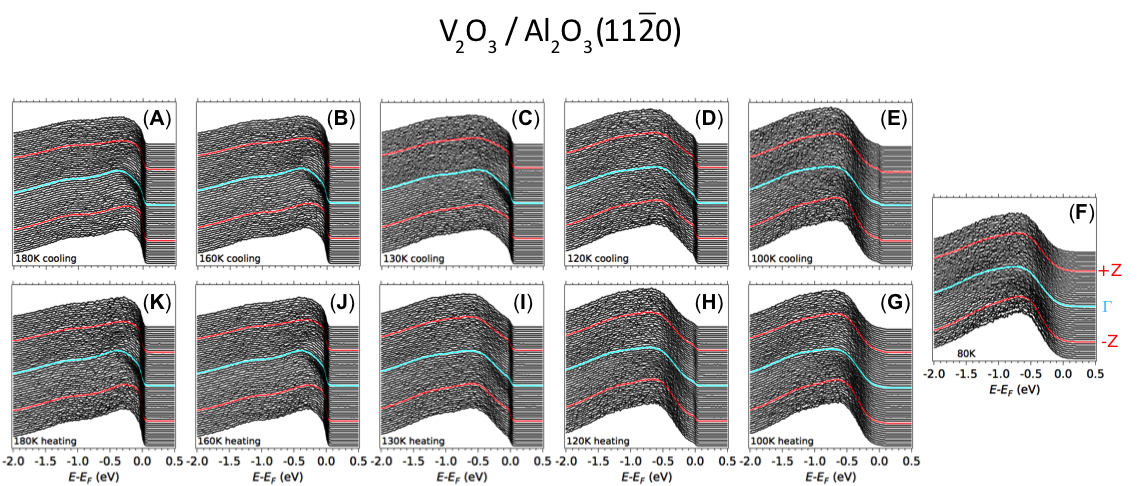}
    \caption{\footnotesize{
        	 \textbf{Complementary temperature-dependent ARPES spectra.}
        	 (A-F) Raw EDC stacks along $\Gamma \textrm{Z}$ ($k_{<111>}$) 
        	 when cooling from $180$~K (metallic state) 
        	 to $60$~K (insulating state) in another V$_2$O$_3$\hkl(11-20) thin film,
        	 different from the one shown in the main text.
        	 (G-K)~Corresponding spectra when heating back to 180K.
        	 The data were measured over an angular range allowing to probe states 
        	 slightly beyond the Brillouin zone edges. EDCs at $\Gamma$ and $\textrm{Z}$
        	 are shown in light blue and red, respectively.
        	 Measurements were performed at $h\nu = 86$~eV using linear horizontal polarization.
        	 }
        }
\label{Fig:V2O3_ARPES_EDCs_July2020}
\end{figure*}

Fig.~\ref{Fig:V2O3_ARPES_EDCs_July2020} presents the raw energy distribution curves (EDCs) 
along $\Gamma \textrm{Z}$ ($k_{<111>}$) of the temperature-dependent data 
in another V$_2$O$_3$\hkl(11-20) thin film, different from the one discussed in the main text.
One clearly sees how the dispersive QP peak at $T \gtrsim 160K$, 
Figs.~\ref{Fig:V2O3_ARPES_EDCs_July2020}(A,~B),
starts loosing spectral weight upon cooling without changing its energy-momentum dispersion
--Figs.~\ref{Fig:V2O3_ARPES_EDCs_July2020}(C,~D).
Simultaneously, the broad QLS at $E - E_F \approx -250$~meV, 
best seen around the $\textrm{Z}$ points, rapidly shifts down in energy
when the temperature drops below about $160$~K.
At $T \lesssim 100K$, Figs.~\ref{Fig:V2O3_ARPES_EDCs_July2020}(E,~F), 
the QP peak has essentially vanished, while the QLS
has dropped to $E - E_F \approx -600$~meV.
Upon heating, the dispersive QP peak re-emerges at a temperature
between $120$~K and $130$~K, while the QLS band shifts up in energy.
The thermal hysteresis in the spectral weight transfer between these two bands
is best seen when comparing, for instance, panels (C) and (I) or panels (D) and (H).

\begin{figure*}[h!t]
	\centering
        \includegraphics[clip, width=0.8\linewidth]{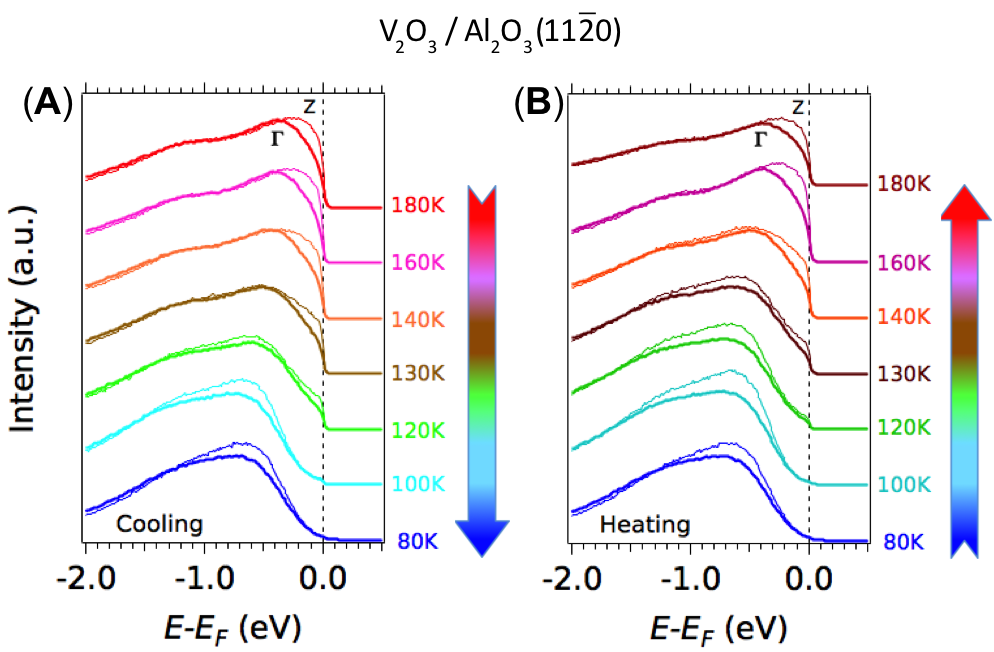}
    \caption{\footnotesize{
          	 \textbf{Complementary temperature-dependent spectra at $\Gamma$ and $+\textrm{Z}$.}
        	 (A,~B)~Energy distribution curves (EDCs) at $\Gamma$ (thick lines) 
        	 and $+\textrm{Z}$ (thin lines), integrated over $0.1$~\AA$^{-1}$ around each point, 
        	 over a cooling and a heating cycle across the MIT, respectively. 
        	 All data in this figure correspond to the spectra shown 
        	 in Fig.~\ref{Fig:V2O3_ARPES_EDCs_July2020} for a V$_2$O$_3$\hkl(11-20) thin film.
        	 Spectra at the same temperature on opposite thermal cycles are shown 
        	 in slightly different color hues, for clarity.
        	 }
        }
\label{Fig:V2O3_EDCs_QP_QLS_July2020}
\end{figure*}

Figures~\ref{Fig:V2O3_EDCs_QP_QLS_July2020}(A,~B) focus on the EDCs 
from Fig.~\ref{Fig:V2O3_ARPES_EDCs_July2020} at the $\Gamma$ and $+\textrm{Z}$ points only.
As the sample is cooled, panel (A), the peak of the QLS at $+\textrm{Z}$ shifts down in energy.
When its energy becomes lower than the bottom of the dispersive QP band, this creates
a downshift of the peak at $\Gamma$, which corresponds to the superposition of the QP band bottom
and the QLS, and a concomitant decrease of its spectral weight 
at energies above the peak position.
Upon heating, panel (B), the opposite behavior, with a clear hysteresis in the peak positions,
is observed.
All the data in this sample thus reproduce the essential observations described in the main text. 


\subsection{Hysteresis in the momentum-integrated ARPES spectra}
\label{Hysteresis_V2O3_aCut_RCut}

\begin{figure}[h!t]
	\centering
        \includegraphics[clip, width=0.9\linewidth]{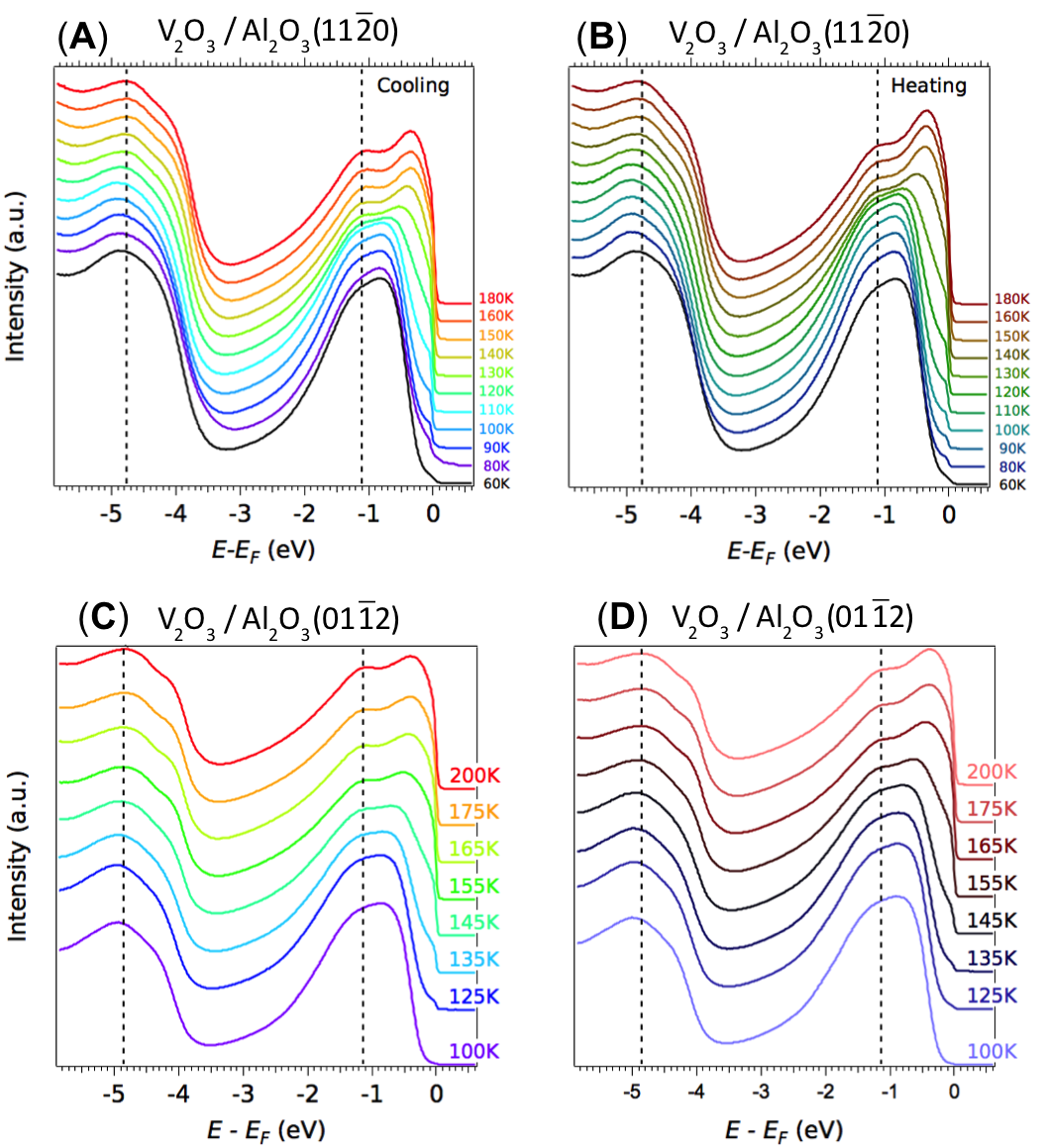}
    \caption{\footnotesize{(Color online)
        	 \textbf{Hysteresis in the momentum-integrated ARPES spectra.}
        	 (A,~B)~Momentum-integrated EDCs on a 
        	 V$_2$O$_3$/Al$_2$O$_3$\hkl(11-20) thin film, 
 			 taken at different temperatures while cooling and heating through the MIT,
 			 respectively. 
			 (C,~D)~Analogous spectra measured on a 
			 V$_2$O$_3$/Al$_2$O$_3$\hkl(01-12) thin film.
			 Dashed lines at $E-E_F \approx -4.8$eV
 			 and $E-E_F \approx -1.1$eV indicate the maxima of the
 			 valence band and MH/OV bands, respectively.
			 Spectra at the same temperature on opposite thermal cycles are shown 
        	 in slightly different color hues, for clarity.
        	 }
        }
\label{Fig:V2O3_PES_hysteresis}
\end{figure}

\begin{figure}[t!h]
	\centering
        \includegraphics[clip, width=0.9\linewidth]{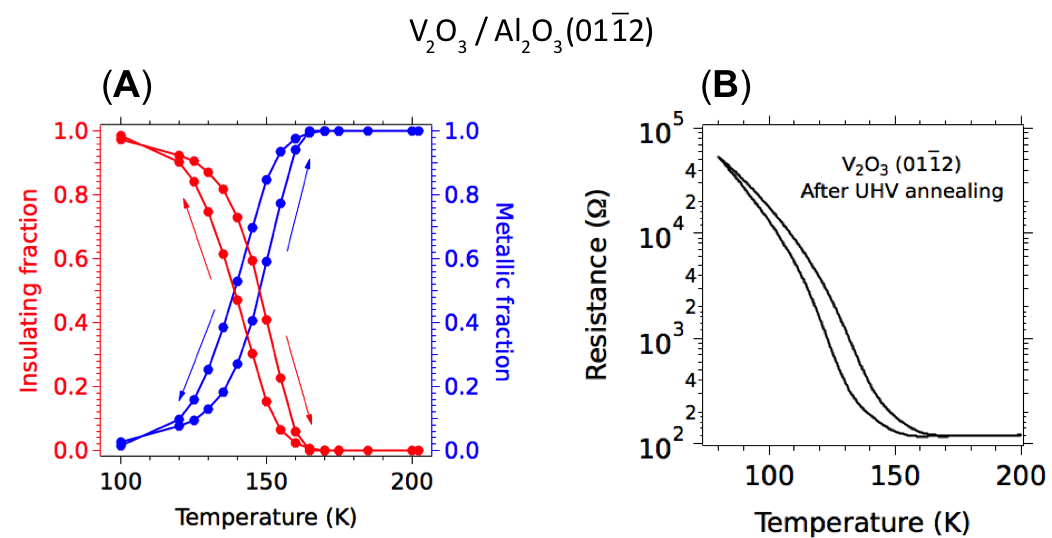}
    \caption{\footnotesize{
			 \textbf{Fraction of metallic and insulating spectral weights 
			 in the ARPES spectra of a V$_2$O$_3$/Al$_2$O$_3$\hkl(01-12) thin film.}
			 (A)~Representation of the fraction of insulating and metallic 
			 domains contributing to the ARPES intensity in the same
			 V$_2$O$_3$/Al$_2$O$_3$\hkl(01-12) thin film of 
			 Figs.~\ref{Fig:V2O3_PES_hysteresis}(C,~D).
			 Error bars in the fits, Eq.~(1) of the main text,
			 are smaller than the size of the symbols.
			 (B)~Resistance as a function of temperature measured in the same
			 sample \emph{after} UHV annealing and photoemission experiments.
        	 }
        }
\label{Fig:V2O3_012_Hyst_vs_R}
\end{figure}

Fig.~\ref{Fig:V2O3_PES_hysteresis} shows the momentum-integrated ARPES spectra 
along $\Gamma \textrm{Z}$ ($k_{<111>}$)
for V$_2$O$_3$/Al$_2$O$_3$\hkl(01-12) and V$_2$O$_3$/Al$_2$O$_3$\hkl(11-20) thin films,
during cooling and heating cycles.
As temperature lowers, the peak near $E_F$ corresponding to the conduction band bundle,
formed by the itinerant QP \emph{and} the quasi-localized states,
shifts down in energy until it reaches an energy $E-E_F \approx -800$meV,
while the spectral weight of the Fermi step at $E_F$ decreases until it ultimately disappears,
thus creating the effect of a gap opening.
The MH/OV band around $E-E_F=-1.1$eV remains at a fixed energy. 
As the system is heated up again, the QP/LS peak shifts back towards $E_F$ 
with a significant hysteresis in temperature, 
characteristic of first-order phase transitions. 
This is especially visible when comparing the EDCs 
between the heating and cooling processes,
around 130K for the V$_2$O$_3$/Al$_2$O$_3$\hkl(01-12) sample,
and around 145K for the V$_2$O$_3$/Al$_2$O$_3$\hkl(11-20) sample.

Finally, Fig.~\ref{Fig:V2O3_012_Hyst_vs_R}(A) shows the temperature-dependent fraction of 
insulating and metallic domains extracted from the ARPES data
of the V$_2$O$_3$/Al$_2$O$_3$\hkl(01-12) sample.
Fig.~\ref{Fig:V2O3_012_Hyst_vs_R}(B) shows, for comparison,
the resistance obtained on the same sample \emph{after} ARPES measurements.

\clearpage


%


\begin{thebibliography}{46}%
\makeatletter
\providecommand \@ifxundefined [1]{%
 \@ifx{#1\undefined}
}%
\providecommand \@ifnum [1]{%
 \ifnum #1\expandafter \@firstoftwo
 \else \expandafter \@secondoftwo
 \fi
}%
\providecommand \@ifx [1]{%
 \ifx #1\expandafter \@firstoftwo
 \else \expandafter \@secondoftwo
 \fi
}%
\providecommand \natexlab [1]{#1}%
\providecommand \enquote  [1]{``#1''}%
\providecommand \bibnamefont  [1]{#1}%
\providecommand \bibfnamefont [1]{#1}%
\providecommand \citenamefont [1]{#1}%
\providecommand \href@noop [0]{\@secondoftwo}%
\providecommand \href [0]{\begingroup \@sanitize@url \@href}%
\providecommand \@href[1]{\@@startlink{#1}\@@href}%
\providecommand \@@href[1]{\endgroup#1\@@endlink}%
\providecommand \@sanitize@url [0]{\catcode `\\12\catcode `\$12\catcode
  `\&12\catcode `\#12\catcode `\^12\catcode `\_12\catcode `\%12\relax}%
\providecommand \@@startlink[1]{}%
\providecommand \@@endlink[0]{}%
\providecommand \url  [0]{\begingroup\@sanitize@url \@url }%
\providecommand \@url [1]{\endgroup\@href {#1}{\urlprefix }}%
\providecommand \urlprefix  [0]{URL }%
\providecommand \Eprint [0]{\href }%
\providecommand \doibase [0]{https://doi.org/}%
\providecommand \selectlanguage [0]{\@gobble}%
\providecommand \bibinfo  [0]{\@secondoftwo}%
\providecommand \bibfield  [0]{\@secondoftwo}%
\providecommand \translation [1]{[#1]}%
\providecommand \BibitemOpen [0]{}%
\providecommand \bibitemStop [0]{}%
\providecommand \bibitemNoStop [0]{.\EOS\space}%
\providecommand \EOS [0]{\spacefactor3000\relax}%
\providecommand \BibitemShut  [1]{\csname bibitem#1\endcsname}%
\let\auto@bib@innerbib\@empty
\bibitem [{\citenamefont {Aschroft}\ and\ \citenamefont
  {Mermin}(1976)}]{Aschroft-Mermin-Book}%
  \BibitemOpen
  \bibfield  {author} {\bibinfo {author} {\bibfnamefont {N.~W.}\ \bibnamefont
  {Aschroft}}\ and\ \bibinfo {author} {\bibfnamefont {N.}~\bibnamefont
  {Mermin}},\ }\href@noop {} {\emph {\bibinfo {title} {Solid State Physics}}}\
  (\bibinfo  {publisher} {Brooks/Cole},\ \bibinfo {year} {1976})\BibitemShut
  {NoStop}%
\bibitem [{\citenamefont {Fulde}(2012)}]{Fulde-2012}%
  \BibitemOpen
  \bibfield  {author} {\bibinfo {author} {\bibfnamefont {P.}~\bibnamefont
  {Fulde}},\ }\href {https://doi.org/10.1142/8419} {\emph {\bibinfo {title}
  {Correlated Electrons in Quantum Matter}}}\ (\bibinfo  {publisher} {World
  Scientific},\ \bibinfo {year} {2012})\ \Eprint
  {https://arxiv.org/abs/https://www.worldscientific.com/doi/pdf/10.1142/8419}
  {https://www.worldscientific.com/doi/pdf/10.1142/8419} \BibitemShut {NoStop}%
\bibitem [{\citenamefont {Mott}(1968)}]{Mott-1968}%
  \BibitemOpen
  \bibfield  {author} {\bibinfo {author} {\bibfnamefont {N.~F.}\ \bibnamefont
  {Mott}},\ }\bibfield  {title} {\bibinfo {title} {Metal-insulator
  transition},\ }\href {https://doi.org/10.1103/RevModPhys.40.677} {\bibfield
  {journal} {\bibinfo  {journal} {Rev. Mod. Phys.}\ }\textbf {\bibinfo {volume}
  {40}},\ \bibinfo {pages} {677} (\bibinfo {year} {1968})}\BibitemShut
  {NoStop}%
\bibitem [{\citenamefont {McWhan}\ \emph {et~al.}(1969)\citenamefont {McWhan},
  \citenamefont {Rice},\ and\ \citenamefont {Remeika}}]{McWhan-1969}%
  \BibitemOpen
  \bibfield  {author} {\bibinfo {author} {\bibfnamefont {D.~B.}\ \bibnamefont
  {McWhan}}, \bibinfo {author} {\bibfnamefont {T.~M.}\ \bibnamefont {Rice}},\
  and\ \bibinfo {author} {\bibfnamefont {J.~P.}\ \bibnamefont {Remeika}},\
  }\bibfield  {title} {\bibinfo {title} {Mott transition in {C}r-doped
  {V}$_2${O}$_3$},\ }\href
  {https://doi.org/https://doi.org/10.1103/PhysRevLett.23.1384} {\bibfield
  {journal} {\bibinfo  {journal} {Phys.~Rev.~Lett}\ }\textbf {\bibinfo {volume}
  {{\bf 23}}},\ \bibinfo {pages} {1384} (\bibinfo {year} {1969})}\BibitemShut
  {NoStop}%
\bibitem [{\citenamefont {McWhan}\ \emph {et~al.}(1971)\citenamefont {McWhan},
  \citenamefont {Remeika}, \citenamefont {Rice}, \citenamefont {Brinkman},
  \citenamefont {Maita},\ and\ \citenamefont {Menth}}]{McWhan-1971}%
  \BibitemOpen
  \bibfield  {author} {\bibinfo {author} {\bibfnamefont {D.~B.}\ \bibnamefont
  {McWhan}}, \bibinfo {author} {\bibfnamefont {J.~P.}\ \bibnamefont {Remeika}},
  \bibinfo {author} {\bibfnamefont {T.~M.}\ \bibnamefont {Rice}}, \bibinfo
  {author} {\bibfnamefont {W.~F.}\ \bibnamefont {Brinkman}}, \bibinfo {author}
  {\bibfnamefont {J.~P.}\ \bibnamefont {Maita}},\ and\ \bibinfo {author}
  {\bibfnamefont {A.}~\bibnamefont {Menth}},\ }\bibfield  {title} {\bibinfo
  {title} {Electronic specific heat of metallic {T}i-doped {V}$_2${O}$_3$},\
  }\href {https://doi.org/10.1103/PhysRevLett.27.941} {\bibfield  {journal}
  {\bibinfo  {journal} {Phys. Rev. Lett.}\ }\textbf {\bibinfo {volume} {27}},\
  \bibinfo {pages} {941} (\bibinfo {year} {1971})}\BibitemShut {NoStop}%
\bibitem [{\citenamefont {Mott}(1990)}]{Mott-1990}%
  \BibitemOpen
  \bibfield  {author} {\bibinfo {author} {\bibfnamefont {N.~F.}\ \bibnamefont
  {Mott}},\ }\href {https://doi.org/https://doi.org/10.1002/crat.2170260620}
  {\emph {\bibinfo {title} {Metal-Insulator Transitions}}}\ (\bibinfo
  {publisher} {Taylor and Francis, London},\ \bibinfo {year}
  {1990})\BibitemShut {NoStop}%
\bibitem [{\citenamefont {Imada}\ \emph {et~al.}(1998)\citenamefont {Imada},
  \citenamefont {Fujimori},\ and\ \citenamefont {Tokura}}]{Imada-1998}%
  \BibitemOpen
  \bibfield  {author} {\bibinfo {author} {\bibfnamefont {M.}~\bibnamefont
  {Imada}}, \bibinfo {author} {\bibfnamefont {A.}~\bibnamefont {Fujimori}},\
  and\ \bibinfo {author} {\bibfnamefont {Y.}~\bibnamefont {Tokura}},\
  }\bibfield  {title} {\bibinfo {title} {Metal-insulator transitions},\ }\href
  {https://doi.org/https://doi.org/10.1103/RevModPhys.70.1039} {\bibfield
  {journal} {\bibinfo  {journal} {Rev.~Mod.~Phys}\ }\textbf {\bibinfo {volume}
  {{\bf 70}}},\ \bibinfo {pages} {1039} (\bibinfo {year} {1998})}\BibitemShut
  {NoStop}%
\bibitem [{\citenamefont {Dobrosavljevic}\ \emph {et~al.}(2012)\citenamefont
  {Dobrosavljevic}, \citenamefont {Trivedi},\ and\ \citenamefont
  {Valles}}]{Dobrosavljevic-2012}%
  \BibitemOpen
  \bibinfo {editor} {\bibfnamefont {V.}~\bibnamefont {Dobrosavljevic}},
  \bibinfo {editor} {\bibfnamefont {N.}~\bibnamefont {Trivedi}},\ and\ \bibinfo
  {editor} {\bibfnamefont {J.~M.}\ \bibnamefont {Valles}, \bibfnamefont
  {Jr.}},\ eds.,\ \href
  {https://doi.org/10.1093/acprof:oso/9780199592593.001.0001} {\emph {\bibinfo
  {title} {Conductor-Insulator Quantum Phase Transitions}}}\ (\bibinfo
  {publisher} {Oxford University},\ \bibinfo {year} {2012})\BibitemShut
  {NoStop}%
\bibitem [{\citenamefont {Rozenberg}\ \emph {et~al.}(1995)\citenamefont
  {Rozenberg}, \citenamefont {Kotliar}, \citenamefont {Kajueter}, \citenamefont
  {Thomas}, \citenamefont {Rapkine}, \citenamefont {Honig},\ and\ \citenamefont
  {Metcalf}}]{Rozenberg-1995}%
  \BibitemOpen
  \bibfield  {author} {\bibinfo {author} {\bibfnamefont {M.~J.}\ \bibnamefont
  {Rozenberg}}, \bibinfo {author} {\bibfnamefont {G.}~\bibnamefont {Kotliar}},
  \bibinfo {author} {\bibfnamefont {H.}~\bibnamefont {Kajueter}}, \bibinfo
  {author} {\bibfnamefont {G.~A.}\ \bibnamefont {Thomas}}, \bibinfo {author}
  {\bibfnamefont {D.~H.}\ \bibnamefont {Rapkine}}, \bibinfo {author}
  {\bibfnamefont {J.~M.}\ \bibnamefont {Honig}},\ and\ \bibinfo {author}
  {\bibfnamefont {P.}~\bibnamefont {Metcalf}},\ }\bibfield  {title} {\bibinfo
  {title} {Optical conductivity in mott-hubbard systems},\ }\href
  {https://doi.org/10.1103/PhysRevLett.75.105} {\bibfield  {journal} {\bibinfo
  {journal} {Phys. Rev. Lett.}\ }\textbf {\bibinfo {volume} {75}},\ \bibinfo
  {pages} {105} (\bibinfo {year} {1995})}\BibitemShut {NoStop}%
\bibitem [{\citenamefont {Rozenberg}\ \emph {et~al.}(1996)\citenamefont
  {Rozenberg}, \citenamefont {Kotliar},\ and\ \citenamefont
  {Kajueter}}]{Rozenberg-1996}%
  \BibitemOpen
  \bibfield  {author} {\bibinfo {author} {\bibfnamefont {M.~J.}\ \bibnamefont
  {Rozenberg}}, \bibinfo {author} {\bibfnamefont {G.}~\bibnamefont {Kotliar}},\
  and\ \bibinfo {author} {\bibfnamefont {H.}~\bibnamefont {Kajueter}},\
  }\bibfield  {title} {\bibinfo {title} {Transfer of spectral weight in
  spectroscopies of correlated electron systems},\ }\href
  {https://doi.org/10.1103/PhysRevB.54.8452} {\bibfield  {journal} {\bibinfo
  {journal} {Phys. Rev. B}\ }\textbf {\bibinfo {volume} {54}},\ \bibinfo
  {pages} {8452} (\bibinfo {year} {1996})}\BibitemShut {NoStop}%
\bibitem [{\citenamefont {Moon}(1970)}]{Moon-1970}%
  \BibitemOpen
  \bibfield  {author} {\bibinfo {author} {\bibfnamefont {R.~M.}\ \bibnamefont
  {Moon}},\ }\bibfield  {title} {\bibinfo {title} {Antiferromagnetism in
  ${\mathrm{v}}_{2}$${\mathrm{o}}_{3}$},\ }\href
  {https://doi.org/10.1103/PhysRevLett.25.527} {\bibfield  {journal} {\bibinfo
  {journal} {Phys. Rev. Lett.}\ }\textbf {\bibinfo {volume} {25}},\ \bibinfo
  {pages} {527} (\bibinfo {year} {1970})}\BibitemShut {NoStop}%
\bibitem [{\citenamefont {Bao}\ \emph {et~al.}(1997)\citenamefont {Bao},
  \citenamefont {Broholm}, \citenamefont {Aeppli}, \citenamefont {Dai},
  \citenamefont {Honig},\ and\ \citenamefont {Metcalf}}]{Bao-1997}%
  \BibitemOpen
  \bibfield  {author} {\bibinfo {author} {\bibfnamefont {W.}~\bibnamefont
  {Bao}}, \bibinfo {author} {\bibfnamefont {C.}~\bibnamefont {Broholm}},
  \bibinfo {author} {\bibfnamefont {G.}~\bibnamefont {Aeppli}}, \bibinfo
  {author} {\bibfnamefont {P.}~\bibnamefont {Dai}}, \bibinfo {author}
  {\bibfnamefont {J.~M.}\ \bibnamefont {Honig}},\ and\ \bibinfo {author}
  {\bibfnamefont {P.}~\bibnamefont {Metcalf}},\ }\bibfield  {title} {\bibinfo
  {title} {Dramatic switching of magnetic exchange in a classic transition
  metal oxide: Evidence for orbital ordering},\ }\href
  {https://doi.org/10.1103/PhysRevLett.78.507} {\bibfield  {journal} {\bibinfo
  {journal} {Phys. Rev. Lett.}\ }\textbf {\bibinfo {volume} {78}},\ \bibinfo
  {pages} {507} (\bibinfo {year} {1997})}\BibitemShut {NoStop}%
\bibitem [{\citenamefont {Anisimov}\ \emph {et~al.}(2005)\citenamefont
  {Anisimov}, \citenamefont {Kondakov}, \citenamefont {Kozhevnikov},
  \citenamefont {Nekrasov}, \citenamefont {Pchelkina}, \citenamefont {Allen},
  \citenamefont {Mo}, \citenamefont {Kim}, \citenamefont {Metcalf},
  \citenamefont {Suga}, \citenamefont {Sekiyama}, \citenamefont {Keller},
  \citenamefont {Leonov}, \citenamefont {Ren},\ and\ \citenamefont
  {Vollhardt}}]{Anisimov-2005}%
  \BibitemOpen
  \bibfield  {author} {\bibinfo {author} {\bibfnamefont {V.~I.}\ \bibnamefont
  {Anisimov}}, \bibinfo {author} {\bibfnamefont {D.~E.}\ \bibnamefont
  {Kondakov}}, \bibinfo {author} {\bibfnamefont {A.~V.}\ \bibnamefont
  {Kozhevnikov}}, \bibinfo {author} {\bibfnamefont {I.~A.}\ \bibnamefont
  {Nekrasov}}, \bibinfo {author} {\bibfnamefont {Z.~V.}\ \bibnamefont
  {Pchelkina}}, \bibinfo {author} {\bibfnamefont {J.~W.}\ \bibnamefont
  {Allen}}, \bibinfo {author} {\bibfnamefont {S.-K.}\ \bibnamefont {Mo}},
  \bibinfo {author} {\bibfnamefont {H.-D.}\ \bibnamefont {Kim}}, \bibinfo
  {author} {\bibfnamefont {P.}~\bibnamefont {Metcalf}}, \bibinfo {author}
  {\bibfnamefont {S.}~\bibnamefont {Suga}}, \bibinfo {author} {\bibfnamefont
  {A.}~\bibnamefont {Sekiyama}}, \bibinfo {author} {\bibfnamefont
  {G.}~\bibnamefont {Keller}}, \bibinfo {author} {\bibfnamefont
  {I.}~\bibnamefont {Leonov}}, \bibinfo {author} {\bibfnamefont
  {X.}~\bibnamefont {Ren}},\ and\ \bibinfo {author} {\bibfnamefont
  {D.}~\bibnamefont {Vollhardt}},\ }\bibfield  {title} {\bibinfo {title} {Full
  orbital calculation scheme for materials with strongly correlated
  electrons},\ }\href
  {https://doi.org/https://doi.org/10.1103/PhysRevB.71.125119} {\bibfield
  {journal} {\bibinfo  {journal} {Phys.~Rev.~B}\ }\textbf {\bibinfo {volume}
  {{\bf 71}}},\ \bibinfo {pages} {125119} (\bibinfo {year} {2005})}\BibitemShut
  {NoStop}%
\bibitem [{\citenamefont {Poteryaev}\ \emph {et~al.}(2007)\citenamefont
  {Poteryaev}, \citenamefont {Tomczak}, \citenamefont {Biermann}, \citenamefont
  {Georges}, \citenamefont {Lichtenstein}, \citenamefont {Rubtsov},
  \citenamefont {Saha-Dasgupta}, ,\ and\ \citenamefont
  {Andersen}}]{Poteryaev-2007}%
  \BibitemOpen
  \bibfield  {author} {\bibinfo {author} {\bibfnamefont {A.~I.}\ \bibnamefont
  {Poteryaev}}, \bibinfo {author} {\bibfnamefont {J.~M.}\ \bibnamefont
  {Tomczak}}, \bibinfo {author} {\bibfnamefont {S.}~\bibnamefont {Biermann}},
  \bibinfo {author} {\bibfnamefont {A.}~\bibnamefont {Georges}}, \bibinfo
  {author} {\bibfnamefont {A.~I.}\ \bibnamefont {Lichtenstein}}, \bibinfo
  {author} {\bibfnamefont {A.~N.}\ \bibnamefont {Rubtsov}}, \bibinfo {author}
  {\bibfnamefont {T.}~\bibnamefont {Saha-Dasgupta}}, ,\ and\ \bibinfo {author}
  {\bibfnamefont {O.~K.}\ \bibnamefont {Andersen}},\ }\bibfield  {title}
  {\bibinfo {title} {Enhanced crystal-field splitting and orbital-selective
  coherence induced by strong correlations in {V}$_2${O}$_3$},\ }\href
  {https://doi.org/https://doi.org/10.1103/PhysRevB.76.085127} {\bibfield
  {journal} {\bibinfo  {journal} {Phys.~Rev.~B}\ }\textbf {\bibinfo {volume}
  {{\bf 76}}},\ \bibinfo {pages} {085127} (\bibinfo {year} {2007})}\BibitemShut
  {NoStop}%
\bibitem [{\citenamefont {Trastoy}\ \emph {et~al.}(2020)\citenamefont
  {Trastoy}, \citenamefont {Camjayi}, \citenamefont {del Valle}, \citenamefont
  {Kalcheim}, \citenamefont {Crocombette}, \citenamefont {Gilbert},
  \citenamefont {Borchers}, \citenamefont {Villegas}, \citenamefont
  {Ravelosona}, \citenamefont {Rozenberg},\ and\ \citenamefont
  {Schuller}}]{Trastoy-2020}%
  \BibitemOpen
  \bibfield  {author} {\bibinfo {author} {\bibfnamefont {J.}~\bibnamefont
  {Trastoy}}, \bibinfo {author} {\bibfnamefont {A.}~\bibnamefont {Camjayi}},
  \bibinfo {author} {\bibfnamefont {J.}~\bibnamefont {del Valle}}, \bibinfo
  {author} {\bibfnamefont {Y.}~\bibnamefont {Kalcheim}}, \bibinfo {author}
  {\bibfnamefont {J.-P.}\ \bibnamefont {Crocombette}}, \bibinfo {author}
  {\bibfnamefont {D.~A.}\ \bibnamefont {Gilbert}}, \bibinfo {author}
  {\bibfnamefont {J.~A.}\ \bibnamefont {Borchers}}, \bibinfo {author}
  {\bibfnamefont {J.~E.}\ \bibnamefont {Villegas}}, \bibinfo {author}
  {\bibfnamefont {D.}~\bibnamefont {Ravelosona}}, \bibinfo {author}
  {\bibfnamefont {M.~J.}\ \bibnamefont {Rozenberg}},\ and\ \bibinfo {author}
  {\bibfnamefont {I.~K.}\ \bibnamefont {Schuller}},\ }\bibfield  {title}
  {\bibinfo {title} {Magnetic field frustration of the metal-insulator
  transition in {V$_2$O$_3$}},\ }\href
  {https://doi.org/10.1103/PhysRevB.101.245109} {\bibfield  {journal} {\bibinfo
   {journal} {Phys. Rev. B}\ }\textbf {\bibinfo {volume} {101}},\ \bibinfo
  {pages} {245109} (\bibinfo {year} {2020})}\BibitemShut {NoStop}%
\bibitem [{\citenamefont {Kalcheim}\ \emph {et~al.}(2020)\citenamefont
  {Kalcheim}, \citenamefont {Camjayi}, \citenamefont {del Valle}, \citenamefont
  {Salev}, \citenamefont {Rozenberg},\ and\ \citenamefont
  {Schuller}}]{Kalcheim-2020}%
  \BibitemOpen
  \bibfield  {author} {\bibinfo {author} {\bibfnamefont {Y.}~\bibnamefont
  {Kalcheim}}, \bibinfo {author} {\bibfnamefont {A.}~\bibnamefont {Camjayi}},
  \bibinfo {author} {\bibfnamefont {J.}~\bibnamefont {del Valle}}, \bibinfo
  {author} {\bibfnamefont {P.}~\bibnamefont {Salev}}, \bibinfo {author}
  {\bibfnamefont {M.}~\bibnamefont {Rozenberg}},\ and\ \bibinfo {author}
  {\bibfnamefont {I.~K.}\ \bibnamefont {Schuller}},\ }\bibfield  {title}
  {\bibinfo {title} {Non-thermal resistive switching in {M}ott insulator
  nanowires},\ }\href {https://doi.org/10.1038/s41467-020-16752-1} {\bibfield
  {journal} {\bibinfo  {journal} {Nat. Commun.}\ }\textbf {\bibinfo {volume}
  {11}},\ \bibinfo {pages} {2985} (\bibinfo {year} {2020})}\BibitemShut
  {NoStop}%
\bibitem [{\citenamefont {Mo}\ \emph {et~al.}(2003)\citenamefont {Mo},
  \citenamefont {Denlinger}, \citenamefont {Kim}, \citenamefont {Park},
  \citenamefont {Allen}, \citenamefont {Sekiyama}, \citenamefont {Yamasaki},
  \citenamefont {Kadono}, \citenamefont {Suga}, \citenamefont {Saitoh},
  \citenamefont {Muro}, \citenamefont {Metcalf}, \citenamefont {Keller},
  \citenamefont {Held}, \citenamefont {Eyert}, \citenamefont {Anisimov},\ and\
  \citenamefont {Vollhardt}}]{Mo-2003}%
  \BibitemOpen
  \bibfield  {author} {\bibinfo {author} {\bibfnamefont {S.-K.}\ \bibnamefont
  {Mo}}, \bibinfo {author} {\bibfnamefont {J.~D.}\ \bibnamefont {Denlinger}},
  \bibinfo {author} {\bibfnamefont {H.-D.}\ \bibnamefont {Kim}}, \bibinfo
  {author} {\bibfnamefont {J.-H.}\ \bibnamefont {Park}}, \bibinfo {author}
  {\bibfnamefont {J.~W.}\ \bibnamefont {Allen}}, \bibinfo {author}
  {\bibfnamefont {A.}~\bibnamefont {Sekiyama}}, \bibinfo {author}
  {\bibfnamefont {A.}~\bibnamefont {Yamasaki}}, \bibinfo {author}
  {\bibfnamefont {K.}~\bibnamefont {Kadono}}, \bibinfo {author} {\bibfnamefont
  {S.}~\bibnamefont {Suga}}, \bibinfo {author} {\bibfnamefont {Y.}~\bibnamefont
  {Saitoh}}, \bibinfo {author} {\bibfnamefont {T.}~\bibnamefont {Muro}},
  \bibinfo {author} {\bibfnamefont {P.}~\bibnamefont {Metcalf}}, \bibinfo
  {author} {\bibfnamefont {G.}~\bibnamefont {Keller}}, \bibinfo {author}
  {\bibfnamefont {K.}~\bibnamefont {Held}}, \bibinfo {author} {\bibfnamefont
  {V.}~\bibnamefont {Eyert}}, \bibinfo {author} {\bibfnamefont {V.~I.}\
  \bibnamefont {Anisimov}},\ and\ \bibinfo {author} {\bibfnamefont
  {D.}~\bibnamefont {Vollhardt}},\ }\bibfield  {title} {\bibinfo {title}
  {Prominent quasiparticle peak in the photoemission spectrum of the metallic
  phase of {V}$_2${O}$_3$},\ }\href
  {https://doi.org/10.1103/PhysRevLett.90.186403} {\bibfield  {journal}
  {\bibinfo  {journal} {Phys. Rev. Lett.}\ }\textbf {\bibinfo {volume} {90}},\
  \bibinfo {pages} {186403} (\bibinfo {year} {2003})}\BibitemShut {NoStop}%
\bibitem [{\citenamefont {Mo}\ \emph {et~al.}(2006)\citenamefont {Mo},
  \citenamefont {Kim}, \citenamefont {Denlinger}, \citenamefont {Allen},
  \citenamefont {Park}, \citenamefont {Sekiyama}, \citenamefont {Yamasaki},
  \citenamefont {Suga}, \citenamefont {Saitoh}, \citenamefont {Muro},\ and\
  \citenamefont {Metcalf}}]{Mo-2006}%
  \BibitemOpen
  \bibfield  {author} {\bibinfo {author} {\bibfnamefont {S.-K.}\ \bibnamefont
  {Mo}}, \bibinfo {author} {\bibfnamefont {H.-D.}\ \bibnamefont {Kim}},
  \bibinfo {author} {\bibfnamefont {J.~D.}\ \bibnamefont {Denlinger}}, \bibinfo
  {author} {\bibfnamefont {J.~W.}\ \bibnamefont {Allen}}, \bibinfo {author}
  {\bibfnamefont {J.-H.}\ \bibnamefont {Park}}, \bibinfo {author}
  {\bibfnamefont {A.}~\bibnamefont {Sekiyama}}, \bibinfo {author}
  {\bibfnamefont {A.}~\bibnamefont {Yamasaki}}, \bibinfo {author}
  {\bibfnamefont {S.}~\bibnamefont {Suga}}, \bibinfo {author} {\bibfnamefont
  {Y.}~\bibnamefont {Saitoh}}, \bibinfo {author} {\bibfnamefont
  {T.}~\bibnamefont {Muro}},\ and\ \bibinfo {author} {\bibfnamefont
  {P.}~\bibnamefont {Metcalf}},\ }\bibfield  {title} {\bibinfo {title}
  {Photoemission study of ({V}$_{1-x}${M}$_x$)$_2${O}$_3$
  $({M}=\mathrm{Cr},\mathrm{Ti})$},\ }\href
  {https://doi.org/10.1103/PhysRevB.74.165101} {\bibfield  {journal} {\bibinfo
  {journal} {Phys. Rev. B}\ }\textbf {\bibinfo {volume} {{\bf 74}}},\ \bibinfo
  {pages} {165101} (\bibinfo {year} {2006})}\BibitemShut {NoStop}%
\bibitem [{\citenamefont {Shin}\ \emph {et~al.}(1990)\citenamefont {Shin},
  \citenamefont {Suga}, \citenamefont {Taniguchi}, \citenamefont {Fujisawa},
  \citenamefont {Kanzaki}, \citenamefont {Fujimori}, \citenamefont {Daimon},
  \citenamefont {Ueda}, \citenamefont {Kosuge},\ and\ \citenamefont
  {Kachi}}]{Shin-1990}%
  \BibitemOpen
  \bibfield  {author} {\bibinfo {author} {\bibfnamefont {S.}~\bibnamefont
  {Shin}}, \bibinfo {author} {\bibfnamefont {S.}~\bibnamefont {Suga}}, \bibinfo
  {author} {\bibfnamefont {M.}~\bibnamefont {Taniguchi}}, \bibinfo {author}
  {\bibfnamefont {M.}~\bibnamefont {Fujisawa}}, \bibinfo {author}
  {\bibfnamefont {H.}~\bibnamefont {Kanzaki}}, \bibinfo {author} {\bibfnamefont
  {A.}~\bibnamefont {Fujimori}}, \bibinfo {author} {\bibfnamefont
  {H.}~\bibnamefont {Daimon}}, \bibinfo {author} {\bibfnamefont
  {Y.}~\bibnamefont {Ueda}}, \bibinfo {author} {\bibfnamefont {K.}~\bibnamefont
  {Kosuge}},\ and\ \bibinfo {author} {\bibfnamefont {S.}~\bibnamefont
  {Kachi}},\ }\bibfield  {title} {\bibinfo {title} {Vacuum-ultraviolet
  reflectance and photoemission study of the metal-insulator phase transitions
  in {VO$_2$}, {V$_6$O$_{13}$}, and {V$_2$O$_3$}},\ }\href
  {https://doi.org/10.1103/PhysRevB.41.4993} {\bibfield  {journal} {\bibinfo
  {journal} {Phys. Rev. B}\ }\textbf {\bibinfo {volume} {41}},\ \bibinfo
  {pages} {4993} (\bibinfo {year} {1990})}\BibitemShut {NoStop}%
\bibitem [{\citenamefont {Smith}\ and\ \citenamefont
  {Henrich}(1994)}]{Smith-1994}%
  \BibitemOpen
  \bibfield  {author} {\bibinfo {author} {\bibfnamefont {K.~E.}\ \bibnamefont
  {Smith}}\ and\ \bibinfo {author} {\bibfnamefont {V.~E.}\ \bibnamefont
  {Henrich}},\ }\bibfield  {title} {\bibinfo {title} {Photoemission study of
  composition- and temperature-induced metal-insulator transitions in
  {Cr}-doped {V$_2$O$_3$}},\ }\href {https://doi.org/10.1103/PhysRevB.50.1382}
  {\bibfield  {journal} {\bibinfo  {journal} {Phys. Rev. B}\ }\textbf {\bibinfo
  {volume} {50}},\ \bibinfo {pages} {1382} (\bibinfo {year}
  {1994})}\BibitemShut {NoStop}%
\bibitem [{\citenamefont {Georges}\ and\ \citenamefont
  {Kotliar}(1992)}]{Georges-1992}%
  \BibitemOpen
  \bibfield  {author} {\bibinfo {author} {\bibfnamefont {A.}~\bibnamefont
  {Georges}}\ and\ \bibinfo {author} {\bibfnamefont {G.}~\bibnamefont
  {Kotliar}},\ }\bibfield  {title} {\bibinfo {title} {Hubbard model in infinite
  dimensions},\ }\href {https://doi.org/10.1103/PhysRevB.45.6479} {\bibfield
  {journal} {\bibinfo  {journal} {Phys. Rev. B}\ }\textbf {\bibinfo {volume}
  {45}},\ \bibinfo {pages} {6479} (\bibinfo {year} {1992})}\BibitemShut
  {NoStop}%
\bibitem [{\citenamefont {Lo~Vecchio}\ \emph {et~al.}(2016)\citenamefont
  {Lo~Vecchio}, \citenamefont {Denlinger}, \citenamefont {Krupin},
  \citenamefont {Kim}, \citenamefont {Metcalf}, \citenamefont {Lupi},
  \citenamefont {Allen},\ and\ \citenamefont {Lanzara}}]{Lo-Vecchio-2016}%
  \BibitemOpen
  \bibfield  {author} {\bibinfo {author} {\bibfnamefont {I.}~\bibnamefont
  {Lo~Vecchio}}, \bibinfo {author} {\bibfnamefont {J.~D.}\ \bibnamefont
  {Denlinger}}, \bibinfo {author} {\bibfnamefont {O.}~\bibnamefont {Krupin}},
  \bibinfo {author} {\bibfnamefont {B.~J.}\ \bibnamefont {Kim}}, \bibinfo
  {author} {\bibfnamefont {P.~A.}\ \bibnamefont {Metcalf}}, \bibinfo {author}
  {\bibfnamefont {S.}~\bibnamefont {Lupi}}, \bibinfo {author} {\bibfnamefont
  {J.~W.}\ \bibnamefont {Allen}},\ and\ \bibinfo {author} {\bibfnamefont
  {A.}~\bibnamefont {Lanzara}},\ }\bibfield  {title} {\bibinfo {title} {Fermi
  surface of metallic {V}$_2${O}$_3$ from angle-resolved photoemission:
  Mid-level filling of ${e}_{g}^{\ensuremath{\pi}}$ bands},\ }\href
  {https://doi.org/10.1103/PhysRevLett.117.166401} {\bibfield  {journal}
  {\bibinfo  {journal} {Phys. Rev. Lett.}\ }\textbf {\bibinfo {volume} {{\bf
  117}}},\ \bibinfo {pages} {166401} (\bibinfo {year} {2016})}\BibitemShut
  {NoStop}%
\bibitem [{\citenamefont {Valmianski}\ \emph {et~al.}(2017)\citenamefont
  {Valmianski}, \citenamefont {Ramirez}, \citenamefont {Urban}, \citenamefont
  {Batlle},\ and\ \citenamefont {Schuller}}]{Valmianski-2017}%
  \BibitemOpen
  \bibfield  {author} {\bibinfo {author} {\bibfnamefont {I.}~\bibnamefont
  {Valmianski}}, \bibinfo {author} {\bibfnamefont {J.~G.}\ \bibnamefont
  {Ramirez}}, \bibinfo {author} {\bibfnamefont {C.}~\bibnamefont {Urban}},
  \bibinfo {author} {\bibfnamefont {X.}~\bibnamefont {Batlle}},\ and\ \bibinfo
  {author} {\bibfnamefont {I.~K.}\ \bibnamefont {Schuller}},\ }\bibfield
  {title} {\bibinfo {title} {Deviation from bulk in the pressure-temperature
  phase diagram of {V}$_2${O}$_3$ thin films},\ }\href
  {https://doi.org/10.1103/PhysRevB.95.155132} {\bibfield  {journal} {\bibinfo
  {journal} {Phys. Rev. B}\ }\textbf {\bibinfo {volume} {{\bf 95}}},\ \bibinfo
  {pages} {155132} (\bibinfo {year} {2017})}\BibitemShut {NoStop}%
\bibitem [{\citenamefont {Trastoy}\ \emph {et~al.}(2018)\citenamefont
  {Trastoy}, \citenamefont {Kalcheim}, \citenamefont {del Valle}, \citenamefont
  {Valmianski},\ and\ \citenamefont {Schuller}}]{Trastoy-2018}%
  \BibitemOpen
  \bibfield  {author} {\bibinfo {author} {\bibfnamefont {J.}~\bibnamefont
  {Trastoy}}, \bibinfo {author} {\bibfnamefont {Y.}~\bibnamefont {Kalcheim}},
  \bibinfo {author} {\bibfnamefont {J.}~\bibnamefont {del Valle}}, \bibinfo
  {author} {\bibfnamefont {I.}~\bibnamefont {Valmianski}},\ and\ \bibinfo
  {author} {\bibfnamefont {I.~K.}\ \bibnamefont {Schuller}},\ }\bibfield
  {title} {\bibinfo {title} {Enhanced metal-insulator transition in
  {V$_2$O$_3$} by thermal quenching after growth},\ }\href
  {https://doi.org/10.1007/s10853-018-2214-7} {\bibfield  {journal} {\bibinfo
  {journal} {J. Mater. Sci.}\ }\textbf {\bibinfo {volume} {53}},\ \bibinfo
  {pages} {9131} (\bibinfo {year} {2018})}\BibitemShut {NoStop}%
\bibitem [{\citenamefont {Kalcheim}\ \emph {et~al.}(2019)\citenamefont
  {Kalcheim}, \citenamefont {Butakov}, \citenamefont {Vargas}, \citenamefont
  {Lee}, \citenamefont {del Valle}, \citenamefont {Trastoy}, \citenamefont
  {Salev}, \citenamefont {Schuller},\ and\ \citenamefont
  {Schuller}}]{Kalcheim-2019}%
  \BibitemOpen
  \bibfield  {author} {\bibinfo {author} {\bibfnamefont {Y.}~\bibnamefont
  {Kalcheim}}, \bibinfo {author} {\bibfnamefont {N.}~\bibnamefont {Butakov}},
  \bibinfo {author} {\bibfnamefont {N.~M.}\ \bibnamefont {Vargas}}, \bibinfo
  {author} {\bibfnamefont {M.-H.}\ \bibnamefont {Lee}}, \bibinfo {author}
  {\bibfnamefont {J.}~\bibnamefont {del Valle}}, \bibinfo {author}
  {\bibfnamefont {J.}~\bibnamefont {Trastoy}}, \bibinfo {author} {\bibfnamefont
  {P.}~\bibnamefont {Salev}}, \bibinfo {author} {\bibfnamefont
  {J.}~\bibnamefont {Schuller}},\ and\ \bibinfo {author} {\bibfnamefont
  {I.~K.}\ \bibnamefont {Schuller}},\ }\bibfield  {title} {\bibinfo {title}
  {Robust coupling between structural and electronic transitions in a {M}ott
  material},\ }\href {https://doi.org/10.1103/PhysRevLett.122.057601}
  {\bibfield  {journal} {\bibinfo  {journal} {Phys. Rev. Lett.}\ }\textbf
  {\bibinfo {volume} {{\bf 122}}},\ \bibinfo {pages} {057601} (\bibinfo {year}
  {2019})}\BibitemShut {NoStop}%
\bibitem [{\citenamefont {R{\"{o}}del}\ \emph {et~al.}(2016)\citenamefont
  {R{\"{o}}del}, \citenamefont {Fortuna}, \citenamefont {Sengupta},
  \citenamefont {Frantzeskakis}, \citenamefont {F\`evre}, \citenamefont
  {Bertran}, \citenamefont {Mercey}, \citenamefont {Matzen}, \citenamefont
  {Agnus}, \citenamefont {Maroutian}, \citenamefont {Lecoeur},\ and\
  \citenamefont {Santander-Syro}}]{Roedel-2016}%
  \BibitemOpen
  \bibfield  {author} {\bibinfo {author} {\bibfnamefont {T.~C.}\ \bibnamefont
  {R{\"{o}}del}}, \bibinfo {author} {\bibfnamefont {F.}~\bibnamefont
  {Fortuna}}, \bibinfo {author} {\bibfnamefont {S.}~\bibnamefont {Sengupta}},
  \bibinfo {author} {\bibfnamefont {E.}~\bibnamefont {Frantzeskakis}}, \bibinfo
  {author} {\bibfnamefont {P.~L.}\ \bibnamefont {F\`evre}}, \bibinfo {author}
  {\bibfnamefont {F.~c.}\ \bibnamefont {Bertran}}, \bibinfo {author}
  {\bibfnamefont {B.}~\bibnamefont {Mercey}}, \bibinfo {author} {\bibfnamefont
  {S.}~\bibnamefont {Matzen}}, \bibinfo {author} {\bibfnamefont
  {G.}~\bibnamefont {Agnus}}, \bibinfo {author} {\bibfnamefont
  {T.}~\bibnamefont {Maroutian}}, \bibinfo {author} {\bibfnamefont
  {P.}~\bibnamefont {Lecoeur}},\ and\ \bibinfo {author} {\bibfnamefont {A.~F.}\
  \bibnamefont {Santander-Syro}},\ }\bibfield  {title} {\bibinfo {title}
  {Universal fabrication of {2D} electron systems in functional oxides},\
  }\href {https://doi.org/10.1002/adma.201505021} {\bibfield  {journal}
  {\bibinfo  {journal} {Advanced Materials}\ }\textbf {\bibinfo {volume} {{\bf
  28}}},\ \bibinfo {pages} {1976} (\bibinfo {year} {2016})}\BibitemShut
  {NoStop}%
\bibitem [{\citenamefont {Backes}\ \emph {et~al.}(2016)\citenamefont {Backes},
  \citenamefont {R{\"{o}}del}, \citenamefont {Fortuna}, \citenamefont
  {Frantzeskakis}, \citenamefont {Le~F\`evre}, \citenamefont {Bertran},
  \citenamefont {Kobayashi}, \citenamefont {Yukawa}, \citenamefont
  {Mitsuhashi}, \citenamefont {Kitamura}, \citenamefont {Horiba}, \citenamefont
  {Kumigashira}, \citenamefont {Saint-Martin}, \citenamefont {Fouchet},
  \citenamefont {Berini}, \citenamefont {Dumont}, \citenamefont {Kim},
  \citenamefont {Lechermann}, \citenamefont {Jeschke}, \citenamefont
  {Rozenberg}, \citenamefont {Valent\'{\i}},\ and\ \citenamefont
  {Santander-Syro}}]{Backes-2016}%
  \BibitemOpen
  \bibfield  {author} {\bibinfo {author} {\bibfnamefont {S.}~\bibnamefont
  {Backes}}, \bibinfo {author} {\bibfnamefont {T.~C.}\ \bibnamefont
  {R{\"{o}}del}}, \bibinfo {author} {\bibfnamefont {F.}~\bibnamefont
  {Fortuna}}, \bibinfo {author} {\bibfnamefont {E.}~\bibnamefont
  {Frantzeskakis}}, \bibinfo {author} {\bibfnamefont {P.}~\bibnamefont
  {Le~F\`evre}}, \bibinfo {author} {\bibfnamefont {F.}~\bibnamefont {Bertran}},
  \bibinfo {author} {\bibfnamefont {M.}~\bibnamefont {Kobayashi}}, \bibinfo
  {author} {\bibfnamefont {R.}~\bibnamefont {Yukawa}}, \bibinfo {author}
  {\bibfnamefont {T.}~\bibnamefont {Mitsuhashi}}, \bibinfo {author}
  {\bibfnamefont {M.}~\bibnamefont {Kitamura}}, \bibinfo {author}
  {\bibfnamefont {K.}~\bibnamefont {Horiba}}, \bibinfo {author} {\bibfnamefont
  {H.}~\bibnamefont {Kumigashira}}, \bibinfo {author} {\bibfnamefont
  {R.}~\bibnamefont {Saint-Martin}}, \bibinfo {author} {\bibfnamefont
  {A.}~\bibnamefont {Fouchet}}, \bibinfo {author} {\bibfnamefont
  {B.}~\bibnamefont {Berini}}, \bibinfo {author} {\bibfnamefont
  {Y.}~\bibnamefont {Dumont}}, \bibinfo {author} {\bibfnamefont {A.~J.}\
  \bibnamefont {Kim}}, \bibinfo {author} {\bibfnamefont {F.}~\bibnamefont
  {Lechermann}}, \bibinfo {author} {\bibfnamefont {H.~O.}\ \bibnamefont
  {Jeschke}}, \bibinfo {author} {\bibfnamefont {M.~J.}\ \bibnamefont
  {Rozenberg}}, \bibinfo {author} {\bibfnamefont {R.}~\bibnamefont
  {Valent\'{\i}}},\ and\ \bibinfo {author} {\bibfnamefont {A.~F.}\ \bibnamefont
  {Santander-Syro}},\ }\bibfield  {title} {\bibinfo {title} {Hubbard band
  versus oxygen vacancy states in the correlated electron metal
  {S}r{V}{O}$_3$},\ }\href {https://doi.org/10.1103/PhysRevB.94.241110}
  {\bibfield  {journal} {\bibinfo  {journal} {Phys. Rev. B}\ }\textbf {\bibinfo
  {volume} {{\bf 94}}},\ \bibinfo {pages} {241110} (\bibinfo {year}
  {2016})}\BibitemShut {NoStop}%
\bibitem [{\citenamefont {R{\"{o}}del}\ \emph {et~al.}(2017)\citenamefont
  {R{\"{o}}del}, \citenamefont {Vivek}, \citenamefont {Fortuna}, \citenamefont
  {Le~F\`evre}, \citenamefont {Bertran}, \citenamefont {Weht}, \citenamefont
  {Goniakowski}, \citenamefont {Gabay},\ and\ \citenamefont
  {Santander-Syro}}]{Roedel-2017}%
  \BibitemOpen
  \bibfield  {author} {\bibinfo {author} {\bibfnamefont {T.~C.}\ \bibnamefont
  {R{\"{o}}del}}, \bibinfo {author} {\bibfnamefont {M.}~\bibnamefont {Vivek}},
  \bibinfo {author} {\bibfnamefont {F.}~\bibnamefont {Fortuna}}, \bibinfo
  {author} {\bibfnamefont {P.}~\bibnamefont {Le~F\`evre}}, \bibinfo {author}
  {\bibfnamefont {F.}~\bibnamefont {Bertran}}, \bibinfo {author} {\bibfnamefont
  {R.}~\bibnamefont {Weht}}, \bibinfo {author} {\bibfnamefont {J.}~\bibnamefont
  {Goniakowski}}, \bibinfo {author} {\bibfnamefont {M.}~\bibnamefont {Gabay}},\
  and\ \bibinfo {author} {\bibfnamefont {A.~F.}\ \bibnamefont
  {Santander-Syro}},\ }\bibfield  {title} {\bibinfo {title} {Two-dimensional
  electron systems in {A}{TiO}$_3$ perovskites ({A}={Ca, Ba, Sr}): Control of
  orbital hybridization and energy order},\ }\href
  {https://doi.org/10.1103/PhysRevB.96.041121} {\bibfield  {journal} {\bibinfo
  {journal} {Phys. Rev. B}\ }\textbf {\bibinfo {volume} {{\bf 96}}},\ \bibinfo
  {pages} {041121} (\bibinfo {year} {2017})}\BibitemShut {NoStop}%
\bibitem [{\citenamefont {L{\"{o}}mker}\ \emph {et~al.}(2017)\citenamefont
  {L{\"{o}}mker}, \citenamefont {R{\"{o}}del}, \citenamefont {Gerber},
  \citenamefont {Fortuna}, \citenamefont {Frantzeskakis}, \citenamefont
  {Le~F\`evre}, \citenamefont {Bertran}, \citenamefont {M{\"{u}}ller},\ and\
  \citenamefont {Santander-Syro}}]{Loemker-2017}%
  \BibitemOpen
  \bibfield  {author} {\bibinfo {author} {\bibfnamefont {P.}~\bibnamefont
  {L{\"{o}}mker}}, \bibinfo {author} {\bibfnamefont {T.~C.}\ \bibnamefont
  {R{\"{o}}del}}, \bibinfo {author} {\bibfnamefont {T.}~\bibnamefont {Gerber}},
  \bibinfo {author} {\bibfnamefont {F.}~\bibnamefont {Fortuna}}, \bibinfo
  {author} {\bibfnamefont {E.}~\bibnamefont {Frantzeskakis}}, \bibinfo {author}
  {\bibfnamefont {P.}~\bibnamefont {Le~F\`evre}}, \bibinfo {author}
  {\bibfnamefont {F.}~\bibnamefont {Bertran}}, \bibinfo {author} {\bibfnamefont
  {M.}~\bibnamefont {M{\"{u}}ller}},\ and\ \bibinfo {author} {\bibfnamefont
  {A.~F.}\ \bibnamefont {Santander-Syro}},\ }\bibfield  {title} {\bibinfo
  {title} {Two-dimensional electron system at the magnetically tunable
  {EuO}/{SrTiO}$_3$ interface},\ }\href
  {https://doi.org/10.1103/PhysRevMaterials.1.062001} {\bibfield  {journal}
  {\bibinfo  {journal} {Phys. Rev. Materials}\ }\textbf {\bibinfo {volume}
  {{\bf 1}}},\ \bibinfo {pages} {062001} (\bibinfo {year} {2017})}\BibitemShut
  {NoStop}%
\bibitem [{\citenamefont {R{\"{o}}del}\ \emph {et~al.}(2018)\citenamefont
  {R{\"{o}}del}, \citenamefont {Dai}, \citenamefont {Fortuna}, \citenamefont
  {Frantzeskakis}, \citenamefont {Le~F\`evre}, \citenamefont {Bertran},
  \citenamefont {Kobayashi}, \citenamefont {Yukawa}, \citenamefont
  {Mitsuhashi}, \citenamefont {Kitamura}, \citenamefont {Horiba}, \citenamefont
  {Kumigashira},\ and\ \citenamefont {Santander-Syro}}]{Roedel-2018}%
  \BibitemOpen
  \bibfield  {author} {\bibinfo {author} {\bibfnamefont {T.~C.}\ \bibnamefont
  {R{\"{o}}del}}, \bibinfo {author} {\bibfnamefont {J.}~\bibnamefont {Dai}},
  \bibinfo {author} {\bibfnamefont {F.}~\bibnamefont {Fortuna}}, \bibinfo
  {author} {\bibfnamefont {E.}~\bibnamefont {Frantzeskakis}}, \bibinfo {author}
  {\bibfnamefont {P.}~\bibnamefont {Le~F\`evre}}, \bibinfo {author}
  {\bibfnamefont {F.}~\bibnamefont {Bertran}}, \bibinfo {author} {\bibfnamefont
  {M.}~\bibnamefont {Kobayashi}}, \bibinfo {author} {\bibfnamefont
  {R.}~\bibnamefont {Yukawa}}, \bibinfo {author} {\bibfnamefont
  {T.}~\bibnamefont {Mitsuhashi}}, \bibinfo {author} {\bibfnamefont
  {M.}~\bibnamefont {Kitamura}}, \bibinfo {author} {\bibfnamefont
  {K.}~\bibnamefont {Horiba}}, \bibinfo {author} {\bibfnamefont
  {H.}~\bibnamefont {Kumigashira}},\ and\ \bibinfo {author} {\bibfnamefont
  {A.~F.}\ \bibnamefont {Santander-Syro}},\ }\bibfield  {title} {\bibinfo
  {title} {High-density two-dimensional electron system induced by oxygen
  vacancies in {ZnO}},\ }\href
  {https://doi.org/10.1103/PhysRevMaterials.2.051601} {\bibfield  {journal}
  {\bibinfo  {journal} {Phys. Rev. Materials}\ }\textbf {\bibinfo {volume}
  {{\bf 2}}},\ \bibinfo {pages} {051601} (\bibinfo {year} {2018})}\BibitemShut
  {NoStop}%
\bibitem [{\citenamefont {Santander-Syro}\ \emph {et~al.}(2011)\citenamefont
  {Santander-Syro}, \citenamefont {Copie}, \citenamefont {Kondo}, \citenamefont
  {Fortuna}, \citenamefont {Pailh\`es}, \citenamefont {Weht}, \citenamefont
  {Qiu}, \citenamefont {Bertran}, \citenamefont {Nicolaou}, \citenamefont
  {Taleb-Ibrahimi}, \citenamefont {F\`evre}, \citenamefont {Herranz},
  \citenamefont {Bibes}, \citenamefont {Reyren}, \citenamefont {Apertet},
  \citenamefont {Lecoeur}, \citenamefont {Barth\'el\'emy},\ and\ \citenamefont
  {Rozenberg}}]{Santander-2011}%
  \BibitemOpen
  \bibfield  {author} {\bibinfo {author} {\bibfnamefont {A.~F.}\ \bibnamefont
  {Santander-Syro}}, \bibinfo {author} {\bibfnamefont {O.}~\bibnamefont
  {Copie}}, \bibinfo {author} {\bibfnamefont {T.}~\bibnamefont {Kondo}},
  \bibinfo {author} {\bibfnamefont {F.}~\bibnamefont {Fortuna}}, \bibinfo
  {author} {\bibfnamefont {S.}~\bibnamefont {Pailh\`es}}, \bibinfo {author}
  {\bibfnamefont {R.}~\bibnamefont {Weht}}, \bibinfo {author} {\bibfnamefont
  {X.~G.}\ \bibnamefont {Qiu}}, \bibinfo {author} {\bibfnamefont
  {F.}~\bibnamefont {Bertran}}, \bibinfo {author} {\bibfnamefont
  {A.}~\bibnamefont {Nicolaou}}, \bibinfo {author} {\bibfnamefont
  {A.}~\bibnamefont {Taleb-Ibrahimi}}, \bibinfo {author} {\bibfnamefont
  {P.~L.}\ \bibnamefont {F\`evre}}, \bibinfo {author} {\bibfnamefont
  {G.}~\bibnamefont {Herranz}}, \bibinfo {author} {\bibfnamefont
  {M.}~\bibnamefont {Bibes}}, \bibinfo {author} {\bibfnamefont
  {N.}~\bibnamefont {Reyren}}, \bibinfo {author} {\bibfnamefont
  {Y.}~\bibnamefont {Apertet}}, \bibinfo {author} {\bibfnamefont
  {P.}~\bibnamefont {Lecoeur}}, \bibinfo {author} {\bibfnamefont
  {A.}~\bibnamefont {Barth\'el\'emy}},\ and\ \bibinfo {author} {\bibfnamefont
  {M.~J.}\ \bibnamefont {Rozenberg}},\ }\bibfield  {title} {\bibinfo {title}
  {Two-dimensional electron gas with universal subbands at the surface of
  {SrTiO$_3$}},\ }\href@noop {} {\bibfield  {journal} {\bibinfo  {journal}
  {Nature}\ }\textbf {\bibinfo {volume} {{\bf 469}}},\ \bibinfo {pages} {189}
  (\bibinfo {year} {2011})}\BibitemShut {NoStop}%
\bibitem [{\citenamefont {Santander-Syro}\ \emph {et~al.}(2012)\citenamefont
  {Santander-Syro}, \citenamefont {Bareille}, \citenamefont {Fortuna},
  \citenamefont {Copie}, \citenamefont {Gabay}, \citenamefont {Bertran},
  \citenamefont {Taleb-Ibrahimi}, \citenamefont {F\`evre}, \citenamefont
  {Herranz}, \citenamefont {Reyren}, \citenamefont {Bibes}, \citenamefont
  {Barth\'el\'emy}, \citenamefont {Lecoeur}, \citenamefont {Guevara},\ and\
  \citenamefont {Rozenberg.}}]{Santander-2012}%
  \BibitemOpen
  \bibfield  {author} {\bibinfo {author} {\bibfnamefont {A.~F.}\ \bibnamefont
  {Santander-Syro}}, \bibinfo {author} {\bibfnamefont {C.}~\bibnamefont
  {Bareille}}, \bibinfo {author} {\bibfnamefont {F.}~\bibnamefont {Fortuna}},
  \bibinfo {author} {\bibfnamefont {O.}~\bibnamefont {Copie}}, \bibinfo
  {author} {\bibfnamefont {M.}~\bibnamefont {Gabay}}, \bibinfo {author}
  {\bibfnamefont {F.}~\bibnamefont {Bertran}}, \bibinfo {author} {\bibfnamefont
  {A.}~\bibnamefont {Taleb-Ibrahimi}}, \bibinfo {author} {\bibfnamefont
  {P.~L.}\ \bibnamefont {F\`evre}}, \bibinfo {author} {\bibfnamefont
  {G.}~\bibnamefont {Herranz}}, \bibinfo {author} {\bibfnamefont
  {N.}~\bibnamefont {Reyren}}, \bibinfo {author} {\bibfnamefont
  {M.}~\bibnamefont {Bibes}}, \bibinfo {author} {\bibfnamefont
  {A.}~\bibnamefont {Barth\'el\'emy}}, \bibinfo {author} {\bibfnamefont
  {P.}~\bibnamefont {Lecoeur}}, \bibinfo {author} {\bibfnamefont
  {J.}~\bibnamefont {Guevara}},\ and\ \bibinfo {author} {\bibfnamefont {M.~J.}\
  \bibnamefont {Rozenberg.}},\ }\bibfield  {title} {\bibinfo {title} {Orbital
  symmetry reconstruction and strong mass renormalization in the
  two-dimensional electron gas at the surface of {KTaO$_3$}},\ }\href@noop {}
  {\bibfield  {journal} {\bibinfo  {journal} {Phys. Rev. B}\ }\textbf {\bibinfo
  {volume} {{\bf 86}}},\ \bibinfo {pages} {121107} (\bibinfo {year}
  {2012})}\BibitemShut {NoStop}%
\bibitem [{\citenamefont {Takizawa}\ \emph {et~al.}(2009)\citenamefont
  {Takizawa}, \citenamefont {Minohara}, \citenamefont {Kumigashira},
  \citenamefont {Toyota}, \citenamefont {Oshima}, \citenamefont {Wadati},
  \citenamefont {Yoshida}, \citenamefont {Fujimori}, \citenamefont {Lippmaa},
  \citenamefont {Kawasaki}, \citenamefont {Koinuma}, \citenamefont {Sordi},\
  and\ \citenamefont {Rozenberg}}]{Takizawa-2009}%
  \BibitemOpen
  \bibfield  {author} {\bibinfo {author} {\bibfnamefont {M.}~\bibnamefont
  {Takizawa}}, \bibinfo {author} {\bibfnamefont {M.}~\bibnamefont {Minohara}},
  \bibinfo {author} {\bibfnamefont {H.}~\bibnamefont {Kumigashira}}, \bibinfo
  {author} {\bibfnamefont {D.}~\bibnamefont {Toyota}}, \bibinfo {author}
  {\bibfnamefont {M.}~\bibnamefont {Oshima}}, \bibinfo {author} {\bibfnamefont
  {H.}~\bibnamefont {Wadati}}, \bibinfo {author} {\bibfnamefont
  {T.}~\bibnamefont {Yoshida}}, \bibinfo {author} {\bibfnamefont
  {A.}~\bibnamefont {Fujimori}}, \bibinfo {author} {\bibfnamefont
  {M.}~\bibnamefont {Lippmaa}}, \bibinfo {author} {\bibfnamefont
  {M.}~\bibnamefont {Kawasaki}}, \bibinfo {author} {\bibfnamefont
  {H.}~\bibnamefont {Koinuma}}, \bibinfo {author} {\bibfnamefont
  {G.}~\bibnamefont {Sordi}},\ and\ \bibinfo {author} {\bibfnamefont
  {M.}~\bibnamefont {Rozenberg}},\ }\bibfield  {title} {\bibinfo {title}
  {Coherent and incoherent $d$ band dispersions in {SrVO}$_3$},\ }\href
  {https://doi.org/10.1103/PhysRevB.80.235104} {\bibfield  {journal} {\bibinfo
  {journal} {Phys. Rev. B}\ }\textbf {\bibinfo {volume} {{\bf 80}}},\ \bibinfo
  {pages} {235104} (\bibinfo {year} {2009})}\BibitemShut {NoStop}%
\bibitem [{\citenamefont {Qazilbash}\ \emph
  {et~al.}(2008{\natexlab{a}})\citenamefont {Qazilbash}, \citenamefont {Brehm},
  \citenamefont {Chae}, \citenamefont {Ho}, \citenamefont {Andreev},
  \citenamefont {Kim}, \citenamefont {Yun}, \citenamefont {Balatsky},
  \citenamefont {Maple}, \citenamefont {Keilmann}, \citenamefont {Kim},\ and\
  \citenamefont {Basov}}]{Qazilbash-2008}%
  \BibitemOpen
  \bibfield  {author} {\bibinfo {author} {\bibfnamefont {M.}~\bibnamefont
  {Qazilbash}}, \bibinfo {author} {\bibfnamefont {M.}~\bibnamefont {Brehm}},
  \bibinfo {author} {\bibfnamefont {B.-G.}\ \bibnamefont {Chae}}, \bibinfo
  {author} {\bibfnamefont {P.-C.}\ \bibnamefont {Ho}}, \bibinfo {author}
  {\bibfnamefont {G.}~\bibnamefont {Andreev}}, \bibinfo {author} {\bibfnamefont
  {B.-J.}\ \bibnamefont {Kim}}, \bibinfo {author} {\bibfnamefont
  {S.}~\bibnamefont {Yun}}, \bibinfo {author} {\bibfnamefont {A.}~\bibnamefont
  {Balatsky}}, \bibinfo {author} {\bibfnamefont {M.}~\bibnamefont {Maple}},
  \bibinfo {author} {\bibfnamefont {F.}~\bibnamefont {Keilmann}}, \bibinfo
  {author} {\bibfnamefont {H.-T.}\ \bibnamefont {Kim}},\ and\ \bibinfo {author}
  {\bibfnamefont {D.}~\bibnamefont {Basov}},\ }\bibfield  {title} {\bibinfo
  {title} {Mott transition in {V}{O}$_2$ revealed by infrared spectroscopy and
  nano-imaging},\ }\href {https://doi.org/10.1126/science.1150124} {\bibfield
  {journal} {\bibinfo  {journal} {Science (New York, N.Y.)}\ }\textbf {\bibinfo
  {volume} {{\bf 318}}},\ \bibinfo {pages} {1750} (\bibinfo {year}
  {2008}{\natexlab{a}})}\BibitemShut {NoStop}%
\bibitem [{\citenamefont {McLeod}\ \emph {et~al.}(2016)\citenamefont {McLeod},
  \citenamefont {Heumen}, \citenamefont {Ramirez}, \citenamefont {Wang},
  \citenamefont {Saerbeck}, \citenamefont {Guenon}, \citenamefont {Goldflam},
  \citenamefont {Anderegg}, \citenamefont {Kelly}, \citenamefont {Mueller},
  \citenamefont {Liu}, \citenamefont {Schuller},\ and\ \citenamefont
  {Basov}}]{McLeod-2016}%
  \BibitemOpen
  \bibfield  {author} {\bibinfo {author} {\bibfnamefont {A.}~\bibnamefont
  {McLeod}}, \bibinfo {author} {\bibfnamefont {E.}~\bibnamefont {Heumen}},
  \bibinfo {author} {\bibfnamefont {J.}~\bibnamefont {Ramirez}}, \bibinfo
  {author} {\bibfnamefont {S.}~\bibnamefont {Wang}}, \bibinfo {author}
  {\bibfnamefont {T.}~\bibnamefont {Saerbeck}}, \bibinfo {author}
  {\bibfnamefont {S.}~\bibnamefont {Guenon}}, \bibinfo {author} {\bibfnamefont
  {M.}~\bibnamefont {Goldflam}}, \bibinfo {author} {\bibfnamefont
  {L.}~\bibnamefont {Anderegg}}, \bibinfo {author} {\bibfnamefont
  {P.}~\bibnamefont {Kelly}}, \bibinfo {author} {\bibfnamefont
  {A.}~\bibnamefont {Mueller}}, \bibinfo {author} {\bibfnamefont
  {M.}~\bibnamefont {Liu}}, \bibinfo {author} {\bibfnamefont {I.}~\bibnamefont
  {Schuller}},\ and\ \bibinfo {author} {\bibfnamefont {D.}~\bibnamefont
  {Basov}},\ }\bibfield  {title} {\bibinfo {title} {Nanotextured phase
  coexistence in the correlated insulator {V}$_2${O}$_3$},\ }\href
  {https://doi.org/10.1038/nphys3882} {\bibfield  {journal} {\bibinfo
  {journal} {Nature Physics}\ }\textbf {\bibinfo {volume} {\bf{13}}},\ \bibinfo
  {pages} {80} (\bibinfo {year} {2016})}\BibitemShut {NoStop}%
\bibitem [{\citenamefont {Lupi}\ \emph {et~al.}(2010)\citenamefont {Lupi},
  \citenamefont {Baldassarre}, \citenamefont {Mansart}, \citenamefont
  {Perucchi}, \citenamefont {Barinov}, \citenamefont {Dudin}, \citenamefont
  {Papalazarou}, \citenamefont {Rodolakis}, \citenamefont {Rueff},
  \citenamefont {Iti\'e}, \citenamefont {Ravy}, \citenamefont {Nicoletti},
  \citenamefont {Postorino}, \citenamefont {Hansmann}, \citenamefont {Parragh},
  \citenamefont {Toschi}, \citenamefont {Saha-Dasgupta}, \citenamefont
  {Andersen}, \citenamefont {Sangiovanni}, \citenamefont {Held},\ and\
  \citenamefont {Marsi}}]{Lupi-2010}%
  \BibitemOpen
  \bibfield  {author} {\bibinfo {author} {\bibfnamefont {S.}~\bibnamefont
  {Lupi}}, \bibinfo {author} {\bibfnamefont {L.}~\bibnamefont {Baldassarre}},
  \bibinfo {author} {\bibfnamefont {B.}~\bibnamefont {Mansart}}, \bibinfo
  {author} {\bibfnamefont {A.}~\bibnamefont {Perucchi}}, \bibinfo {author}
  {\bibfnamefont {A.}~\bibnamefont {Barinov}}, \bibinfo {author} {\bibfnamefont
  {P.}~\bibnamefont {Dudin}}, \bibinfo {author} {\bibfnamefont
  {E.}~\bibnamefont {Papalazarou}}, \bibinfo {author} {\bibfnamefont
  {F.}~\bibnamefont {Rodolakis}}, \bibinfo {author} {\bibfnamefont {J.~P.}\
  \bibnamefont {Rueff}}, \bibinfo {author} {\bibfnamefont {J.~P.}\ \bibnamefont
  {Iti\'e}}, \bibinfo {author} {\bibfnamefont {S.}~\bibnamefont {Ravy}},
  \bibinfo {author} {\bibfnamefont {D.}~\bibnamefont {Nicoletti}}, \bibinfo
  {author} {\bibfnamefont {P.}~\bibnamefont {Postorino}}, \bibinfo {author}
  {\bibfnamefont {P.}~\bibnamefont {Hansmann}}, \bibinfo {author}
  {\bibfnamefont {N.}~\bibnamefont {Parragh}}, \bibinfo {author} {\bibfnamefont
  {A.}~\bibnamefont {Toschi}}, \bibinfo {author} {\bibfnamefont
  {T.}~\bibnamefont {Saha-Dasgupta}}, \bibinfo {author} {\bibfnamefont {O.~K.}\
  \bibnamefont {Andersen}}, \bibinfo {author} {\bibfnamefont {G.}~\bibnamefont
  {Sangiovanni}}, \bibinfo {author} {\bibfnamefont {K.}~\bibnamefont {Held}},\
  and\ \bibinfo {author} {\bibfnamefont {M.}~\bibnamefont {Marsi}},\ }\bibfield
   {title} {\bibinfo {title} {A microscopic view on the {M}ott transition in
  chromium-doped {V}$_2${O}$_3$},\ }\href@noop {} {\bibfield  {journal}
  {\bibinfo  {journal} {Nat. Commun.}\ }\textbf {\bibinfo {volume} {{\bf 1}}},\
  \bibinfo {pages} {105} (\bibinfo {year} {2010})}\BibitemShut {NoStop}%
\bibitem [{\citenamefont {Frandsen}\ \emph {et~al.}(2016)\citenamefont
  {Frandsen}, \citenamefont {Liu}, \citenamefont {Cheung}, \citenamefont
  {Guguchia}, \citenamefont {Khasanov}, \citenamefont {Morenzoni},
  \citenamefont {Munsie}, \citenamefont {Hallas}, \citenamefont {Wilson},
  \citenamefont {Cai}, \citenamefont {Luke}, \citenamefont {Chen},
  \citenamefont {Li}, \citenamefont {Jin}, \citenamefont {Ding}, \citenamefont
  {Guo}, \citenamefont {Ning}, \citenamefont {Ito}, \citenamefont {Higemoto},
  \citenamefont {Billinge}, \citenamefont {Sakamoto}, \citenamefont {Fujimori},
  \citenamefont {Murakami}, \citenamefont {Kageyama}, \citenamefont {Alonso},
  \citenamefont {Kotliar}, \citenamefont {Imada},\ and\ \citenamefont
  {Uemura}}]{Frandsen-2016}%
  \BibitemOpen
  \bibfield  {author} {\bibinfo {author} {\bibfnamefont {B.~A.}\ \bibnamefont
  {Frandsen}}, \bibinfo {author} {\bibfnamefont {L.}~\bibnamefont {Liu}},
  \bibinfo {author} {\bibfnamefont {S.~C.}\ \bibnamefont {Cheung}}, \bibinfo
  {author} {\bibfnamefont {Z.}~\bibnamefont {Guguchia}}, \bibinfo {author}
  {\bibfnamefont {R.}~\bibnamefont {Khasanov}}, \bibinfo {author}
  {\bibfnamefont {E.}~\bibnamefont {Morenzoni}}, \bibinfo {author}
  {\bibfnamefont {T.~J.~S.}\ \bibnamefont {Munsie}}, \bibinfo {author}
  {\bibfnamefont {A.~M.}\ \bibnamefont {Hallas}}, \bibinfo {author}
  {\bibfnamefont {M.~N.}\ \bibnamefont {Wilson}}, \bibinfo {author}
  {\bibfnamefont {Y.}~\bibnamefont {Cai}}, \bibinfo {author} {\bibfnamefont
  {G.~M.}\ \bibnamefont {Luke}}, \bibinfo {author} {\bibfnamefont
  {B.}~\bibnamefont {Chen}}, \bibinfo {author} {\bibfnamefont {W.}~\bibnamefont
  {Li}}, \bibinfo {author} {\bibfnamefont {C.}~\bibnamefont {Jin}}, \bibinfo
  {author} {\bibfnamefont {C.}~\bibnamefont {Ding}}, \bibinfo {author}
  {\bibfnamefont {S.}~\bibnamefont {Guo}}, \bibinfo {author} {\bibfnamefont
  {F.}~\bibnamefont {Ning}}, \bibinfo {author} {\bibfnamefont {T.~U.}\
  \bibnamefont {Ito}}, \bibinfo {author} {\bibfnamefont {W.}~\bibnamefont
  {Higemoto}}, \bibinfo {author} {\bibfnamefont {S.~J.~L.}\ \bibnamefont
  {Billinge}}, \bibinfo {author} {\bibfnamefont {S.}~\bibnamefont {Sakamoto}},
  \bibinfo {author} {\bibfnamefont {A.}~\bibnamefont {Fujimori}}, \bibinfo
  {author} {\bibfnamefont {T.}~\bibnamefont {Murakami}}, \bibinfo {author}
  {\bibfnamefont {H.}~\bibnamefont {Kageyama}}, \bibinfo {author}
  {\bibfnamefont {J.~A.}\ \bibnamefont {Alonso}}, \bibinfo {author}
  {\bibfnamefont {G.}~\bibnamefont {Kotliar}}, \bibinfo {author} {\bibfnamefont
  {M.}~\bibnamefont {Imada}},\ and\ \bibinfo {author} {\bibfnamefont {Y.~J.}\
  \bibnamefont {Uemura}},\ }\bibfield  {title} {\bibinfo {title} {Volume-wise
  destruction of the antiferromagnetic {M}ott insulating state through quantum
  tuning},\ }\href {https://doi.org/10.1038/ncomms12519} {\bibfield  {journal}
  {\bibinfo  {journal} {Nature Communications}\ }\textbf {\bibinfo {volume}
  {7}},\ \bibinfo {pages} {12519} (\bibinfo {year} {2016})}\BibitemShut
  {NoStop}%
\bibitem [{\citenamefont {Qazilbash}\ \emph
  {et~al.}(2008{\natexlab{b}})\citenamefont {Qazilbash}, \citenamefont
  {Schafgans}, \citenamefont {Burch}, \citenamefont {Yun}, \citenamefont
  {Chae}, \citenamefont {Kim}, \citenamefont {Kim},\ and\ \citenamefont
  {Basov}}]{Qazilbash-PRB-2008}%
  \BibitemOpen
  \bibfield  {author} {\bibinfo {author} {\bibfnamefont {M.~M.}\ \bibnamefont
  {Qazilbash}}, \bibinfo {author} {\bibfnamefont {A.~A.}\ \bibnamefont
  {Schafgans}}, \bibinfo {author} {\bibfnamefont {K.~S.}\ \bibnamefont
  {Burch}}, \bibinfo {author} {\bibfnamefont {S.~J.}\ \bibnamefont {Yun}},
  \bibinfo {author} {\bibfnamefont {B.~G.}\ \bibnamefont {Chae}}, \bibinfo
  {author} {\bibfnamefont {B.~J.}\ \bibnamefont {Kim}}, \bibinfo {author}
  {\bibfnamefont {H.~T.}\ \bibnamefont {Kim}},\ and\ \bibinfo {author}
  {\bibfnamefont {D.~N.}\ \bibnamefont {Basov}},\ }\bibfield  {title} {\bibinfo
  {title} {Electrodynamics of the vanadium oxides $\mathrm{V}{\mathrm{o}}_{2}$
  and ${\mathrm{v}}_{2}{\mathrm{o}}_{3}$},\ }\href
  {https://doi.org/10.1103/PhysRevB.77.115121} {\bibfield  {journal} {\bibinfo
  {journal} {Phys. Rev. B}\ }\textbf {\bibinfo {volume} {77}},\ \bibinfo
  {pages} {115121} (\bibinfo {year} {2008}{\natexlab{b}})}\BibitemShut
  {NoStop}%
\bibitem [{\citenamefont {Stewart}\ \emph {et~al.}(2012)\citenamefont
  {Stewart}, \citenamefont {Brownstead}, \citenamefont {Wang}, \citenamefont
  {West}, \citenamefont {Ramirez}, \citenamefont {Qazilbash}, \citenamefont
  {Perkins}, \citenamefont {Schuller},\ and\ \citenamefont
  {Basov}}]{Stewart-2012}%
  \BibitemOpen
  \bibfield  {author} {\bibinfo {author} {\bibfnamefont {M.~K.}\ \bibnamefont
  {Stewart}}, \bibinfo {author} {\bibfnamefont {D.}~\bibnamefont {Brownstead}},
  \bibinfo {author} {\bibfnamefont {S.}~\bibnamefont {Wang}}, \bibinfo {author}
  {\bibfnamefont {K.~G.}\ \bibnamefont {West}}, \bibinfo {author}
  {\bibfnamefont {J.~G.}\ \bibnamefont {Ramirez}}, \bibinfo {author}
  {\bibfnamefont {M.~M.}\ \bibnamefont {Qazilbash}}, \bibinfo {author}
  {\bibfnamefont {N.~B.}\ \bibnamefont {Perkins}}, \bibinfo {author}
  {\bibfnamefont {I.~K.}\ \bibnamefont {Schuller}},\ and\ \bibinfo {author}
  {\bibfnamefont {D.~N.}\ \bibnamefont {Basov}},\ }\bibfield  {title} {\bibinfo
  {title} {Insulator-to-metal transition and correlated metallic state of
  {V$_2$O$_3$} investigated by optical spectroscopy},\ }\href
  {https://doi.org/10.1103/PhysRevB.85.205113} {\bibfield  {journal} {\bibinfo
  {journal} {Phys. Rev. B}\ }\textbf {\bibinfo {volume} {85}},\ \bibinfo
  {pages} {205113} (\bibinfo {year} {2012})}\BibitemShut {NoStop}%
\bibitem [{\citenamefont {Lo~Vecchio}\ \emph {et~al.}(2015)\citenamefont
  {Lo~Vecchio}, \citenamefont {Baldassarre}, \citenamefont {D'Apuzzo},
  \citenamefont {Limaj}, \citenamefont {Nicoletti}, \citenamefont {Perucchi},
  \citenamefont {Fan}, \citenamefont {Metcalf}, \citenamefont {Marsi},\ and\
  \citenamefont {Lupi}}]{Lo-Vecchio-2015}%
  \BibitemOpen
  \bibfield  {author} {\bibinfo {author} {\bibfnamefont {I.}~\bibnamefont
  {Lo~Vecchio}}, \bibinfo {author} {\bibfnamefont {L.}~\bibnamefont
  {Baldassarre}}, \bibinfo {author} {\bibfnamefont {F.}~\bibnamefont
  {D'Apuzzo}}, \bibinfo {author} {\bibfnamefont {O.}~\bibnamefont {Limaj}},
  \bibinfo {author} {\bibfnamefont {D.}~\bibnamefont {Nicoletti}}, \bibinfo
  {author} {\bibfnamefont {A.}~\bibnamefont {Perucchi}}, \bibinfo {author}
  {\bibfnamefont {L.}~\bibnamefont {Fan}}, \bibinfo {author} {\bibfnamefont
  {P.}~\bibnamefont {Metcalf}}, \bibinfo {author} {\bibfnamefont
  {M.}~\bibnamefont {Marsi}},\ and\ \bibinfo {author} {\bibfnamefont
  {S.}~\bibnamefont {Lupi}},\ }\bibfield  {title} {\bibinfo {title} {Optical
  properties of ${\mathrm{v}}_{2}{\mathrm{o}}_{3}$ in its whole phase
  diagram},\ }\href {https://doi.org/10.1103/PhysRevB.91.155133} {\bibfield
  {journal} {\bibinfo  {journal} {Phys. Rev. B}\ }\textbf {\bibinfo {volume}
  {91}},\ \bibinfo {pages} {155133} (\bibinfo {year} {2015})}\BibitemShut
  {NoStop}%
\bibitem [{\citenamefont {Park}\ \emph {et~al.}(2000)\citenamefont {Park},
  \citenamefont {Tjeng}, \citenamefont {Tanaka}, \citenamefont {Allen},
  \citenamefont {Chen}, \citenamefont {Metcalf}, \citenamefont {Honig},
  \citenamefont {de~Groot},\ and\ \citenamefont {Sawatzky}}]{Park-2000}%
  \BibitemOpen
  \bibfield  {author} {\bibinfo {author} {\bibfnamefont {J.-H.}\ \bibnamefont
  {Park}}, \bibinfo {author} {\bibfnamefont {L.~H.}\ \bibnamefont {Tjeng}},
  \bibinfo {author} {\bibfnamefont {A.}~\bibnamefont {Tanaka}}, \bibinfo
  {author} {\bibfnamefont {J.~W.}\ \bibnamefont {Allen}}, \bibinfo {author}
  {\bibfnamefont {C.~T.}\ \bibnamefont {Chen}}, \bibinfo {author}
  {\bibfnamefont {P.}~\bibnamefont {Metcalf}}, \bibinfo {author} {\bibfnamefont
  {J.~M.}\ \bibnamefont {Honig}}, \bibinfo {author} {\bibfnamefont {F.~M.~F.}\
  \bibnamefont {de~Groot}},\ and\ \bibinfo {author} {\bibfnamefont {G.~A.}\
  \bibnamefont {Sawatzky}},\ }\bibfield  {title} {\bibinfo {title} {Spin and
  orbital occupation and phase transitions in {V}$_2${O}$_3$},\ }\href
  {https://doi.org/10.1103/PhysRevB.61.11506} {\bibfield  {journal} {\bibinfo
  {journal} {Phys. Rev. B}\ }\textbf {\bibinfo {volume} {61}},\ \bibinfo
  {pages} {11506} (\bibinfo {year} {2000})}\BibitemShut {NoStop}%
\bibitem [{\citenamefont {Rodolakis}\ \emph {et~al.}(2010)\citenamefont
  {Rodolakis}, \citenamefont {Hansmann}, \citenamefont {Rueff}, \citenamefont
  {Toschi}, \citenamefont {Haverkort}, \citenamefont {Sangiovanni},
  \citenamefont {Tanaka}, \citenamefont {Saha-Dasgupta}, \citenamefont
  {Andersen}, \citenamefont {Held}, \citenamefont {Sikora}, \citenamefont
  {Alliot}, \citenamefont {Iti\'e}, \citenamefont {Baudelet}, \citenamefont
  {Wzietek}, \citenamefont {Metcalf},\ and\ \citenamefont
  {Marsi}}]{Rodolakis-2010}%
  \BibitemOpen
  \bibfield  {author} {\bibinfo {author} {\bibfnamefont {F.}~\bibnamefont
  {Rodolakis}}, \bibinfo {author} {\bibfnamefont {P.}~\bibnamefont {Hansmann}},
  \bibinfo {author} {\bibfnamefont {J.-P.}\ \bibnamefont {Rueff}}, \bibinfo
  {author} {\bibfnamefont {A.}~\bibnamefont {Toschi}}, \bibinfo {author}
  {\bibfnamefont {M.~W.}\ \bibnamefont {Haverkort}}, \bibinfo {author}
  {\bibfnamefont {G.}~\bibnamefont {Sangiovanni}}, \bibinfo {author}
  {\bibfnamefont {A.}~\bibnamefont {Tanaka}}, \bibinfo {author} {\bibfnamefont
  {T.}~\bibnamefont {Saha-Dasgupta}}, \bibinfo {author} {\bibfnamefont {O.~K.}\
  \bibnamefont {Andersen}}, \bibinfo {author} {\bibfnamefont {K.}~\bibnamefont
  {Held}}, \bibinfo {author} {\bibfnamefont {M.}~\bibnamefont {Sikora}},
  \bibinfo {author} {\bibfnamefont {I.}~\bibnamefont {Alliot}}, \bibinfo
  {author} {\bibfnamefont {J.-P.}\ \bibnamefont {Iti\'e}}, \bibinfo {author}
  {\bibfnamefont {F.}~\bibnamefont {Baudelet}}, \bibinfo {author}
  {\bibfnamefont {P.}~\bibnamefont {Wzietek}}, \bibinfo {author} {\bibfnamefont
  {P.}~\bibnamefont {Metcalf}},\ and\ \bibinfo {author} {\bibfnamefont
  {M.}~\bibnamefont {Marsi}},\ }\bibfield  {title} {\bibinfo {title}
  {Inequivalent routes across the {M}ott transition in {V$_2$O$_3$} explored by
  x-ray absorption},\ }\href {https://doi.org/10.1103/PhysRevLett.104.047401}
  {\bibfield  {journal} {\bibinfo  {journal} {Phys. Rev. Lett.}\ }\textbf
  {\bibinfo {volume} {104}},\ \bibinfo {pages} {047401} (\bibinfo {year}
  {2010})}\BibitemShut {NoStop}%
\bibitem [{\citenamefont {Grieger}\ and\ \citenamefont
  {Fabrizio}(2015)}]{Grieger-2015}%
  \BibitemOpen
  \bibfield  {author} {\bibinfo {author} {\bibfnamefont {D.}~\bibnamefont
  {Grieger}}\ and\ \bibinfo {author} {\bibfnamefont {M.}~\bibnamefont
  {Fabrizio}},\ }\bibfield  {title} {\bibinfo {title} {Low-temperature magnetic
  ordering and structural distortions in vanadium sesquioxide {V}$_2${O}$_3$},\
  }\href {https://doi.org/10.1103/PhysRevB.92.075121} {\bibfield  {journal}
  {\bibinfo  {journal} {Phys. Rev. B}\ }\textbf {\bibinfo {volume} {92}},\
  \bibinfo {pages} {075121} (\bibinfo {year} {2015})}\BibitemShut {NoStop}%
\bibitem [{\citenamefont {Lechermann}\ \emph {et~al.}(2018)\citenamefont
  {Lechermann}, \citenamefont {Bernstein}, \citenamefont {Mazin},\ and\
  \citenamefont {Valent\'{\i}}}]{Lechermann-2018}%
  \BibitemOpen
  \bibfield  {author} {\bibinfo {author} {\bibfnamefont {F.}~\bibnamefont
  {Lechermann}}, \bibinfo {author} {\bibfnamefont {N.}~\bibnamefont
  {Bernstein}}, \bibinfo {author} {\bibfnamefont {I.~I.}\ \bibnamefont
  {Mazin}},\ and\ \bibinfo {author} {\bibfnamefont {R.}~\bibnamefont
  {Valent\'{\i}}},\ }\bibfield  {title} {\bibinfo {title} {Uncovering the
  mechanism of the impurity-selective {M}ott transition in paramagnetic
  {V}$_2${O}$_3$},\ }\href {https://doi.org/10.1103/PhysRevLett.121.106401}
  {\bibfield  {journal} {\bibinfo  {journal} {Phys. Rev. Lett.}\ }\textbf
  {\bibinfo {volume} {121}},\ \bibinfo {pages} {106401} (\bibinfo {year}
  {2018})}\BibitemShut {NoStop}%
\bibitem [{\citenamefont {Poteryaev}\ \emph {et~al.}(2008)\citenamefont
  {Poteryaev}, \citenamefont {Ferrero}, \citenamefont {Georges},\ and\
  \citenamefont {Parcollet}}]{Poteryaev-2008}%
  \BibitemOpen
  \bibfield  {author} {\bibinfo {author} {\bibfnamefont {A.~I.}\ \bibnamefont
  {Poteryaev}}, \bibinfo {author} {\bibfnamefont {M.}~\bibnamefont {Ferrero}},
  \bibinfo {author} {\bibfnamefont {A.}~\bibnamefont {Georges}},\ and\ \bibinfo
  {author} {\bibfnamefont {O.}~\bibnamefont {Parcollet}},\ }\bibfield  {title}
  {\bibinfo {title} {Effect of crystal-field splitting and interband
  hybridization on the metal-insulator transitions of strongly correlated
  systems},\ }\href {https://doi.org/10.1103/PhysRevB.78.045115} {\bibfield
  {journal} {\bibinfo  {journal} {Phys. Rev. B}\ }\textbf {\bibinfo {volume}
  {78}},\ \bibinfo {pages} {045115} (\bibinfo {year} {2008})}\BibitemShut
  {NoStop}%
\bibitem [{\citenamefont {Zhang}\ \emph {et~al.}(2011)\citenamefont {Zhang},
  \citenamefont {Richard}, \citenamefont {Qian}, \citenamefont {Xu},
  \citenamefont {Dai},\ and\ \citenamefont {Ding}}]{Zhang2011}%
  \BibitemOpen
  \bibfield  {author} {\bibinfo {author} {\bibfnamefont {P.}~\bibnamefont
  {Zhang}}, \bibinfo {author} {\bibfnamefont {P.}~\bibnamefont {Richard}},
  \bibinfo {author} {\bibfnamefont {T.}~\bibnamefont {Qian}}, \bibinfo {author}
  {\bibfnamefont {Y.-M.}\ \bibnamefont {Xu}}, \bibinfo {author} {\bibfnamefont
  {X.}~\bibnamefont {Dai}},\ and\ \bibinfo {author} {\bibfnamefont
  {H.}~\bibnamefont {Ding}},\ }\bibfield  {title} {\bibinfo {title} {A precise
  method for visualizing dispersive features in image plots},\ }\href
  {https://doi.org/10.1063/1.3585113} {\bibfield  {journal} {\bibinfo
  {journal} {Rev. Sci. Instr.}\ }\textbf {\bibinfo {volume} {82}},\ \bibinfo
  {pages} {043712} (\bibinfo {year} {2011})}\BibitemShut {NoStop}%
\end{thebibliography}
\end{document}